\documentclass{iopjournal}

\usepackage{ragged2e}
\usepackage{booktabs}
\usepackage{amsmath}
\usepackage{xcolor}
\usepackage[numbers]{natbib}
\usepackage[T1]{fontenc}
\usepackage[utf8]{inputenc}
\AtBeginDocument{\justifying\setlength{\emergencystretch}{2em}}

\begin{document}

\articletype{Topical Review}

\title{Neutrino Experiments at the LHC and the HL-LHC}

\author{Chayanit Asawatangtrakuldee$^{1,*}$ and Albert De Roeck$^{1,2}$}

\affil{$^{1,*}$ Department of Physics, Faculty of Science, Chulalongkorn University, Bangkok, 10330, Thailand}

\affil{$^2$ High Energy Physics Group, Blackett Laboratory, Imperial College, Prince Consort Road, London, SW7 2AZ, United Kingdom}

\email{chayanit@cern.ch}

\keywords{Far-forward region, TeV, Neutrino, Collider, LHC, HL-LHC}

\begin{abstract}
High energy neutrinos are copiously produced in proton--proton collisions at the Large Hadron Collider (LHC), located at CERN, near Geneva, Switzerland. In particular, a large neutrino flux is expected in the far-forward direction. This flux is concentrated near the beam axis, reaches energies up to the TeV scale, and for electron and tau neutrinos can
be dominated by heavy flavour decays at the highest energies. These features open a new experimental window for studying neutrino interactions with dedicated forward detectors, in an energy range beyond the reach of traditional accelerator-based neutrino experiments. The new FASER and SND@LHC experiments reported the first direct observations of neutrino interactions at a hadron collider using the first data from Run~3, which started in 2022. Since then, the programme has advanced to the first measurements, including charged-current neutrino--nucleon cross sections and the detection of different neutrino flavours. This article describes the forward neutrino flux at the LHC, the acceptance and detector concepts of the Run~3 experiments, and summarises the current status of collider neutrino measurements. It also discusses prospects for extending this neutrino programme towards the HL-LHC era. The collected data will also provide constraints on very forward hadron production, and enable searches for feebly interacting particles.
\end{abstract}

\section{Introduction}
\label{sec:intro}
Neutrino scattering is a direct probe of the weak interaction. It provides precision measurements of the Standard Model~\cite{Brock:1993sz,Conrad:1997ne,Formaggio:2013kya,DeLellis:2004sg} and is also sensitive to physics beyond the Standard Model (BSM)~\cite{Marfatia:2015hja,Arguelles:2019ziu}. However, neutrino interactions at a hadron collider were long assumed to be out of experimental reach.

Neutrinos are copiously produced in proton--proton ($pp$) collisions at the Large Hadron Collider (LHC), but a large fraction of these neutrinos travel in a so-called ``forward beam direction'' and are therefore strongly collimated around the beam collision axis. This neutrino flux contains, at high energies, a large component from decays of heavy flavour hadrons -- charm and bottom. Electroweak sources such as $W/Z$ production become increasingly relevant at more central angles in the detectors, away from the beam axis.

The collider neutrino beam kinematics motivate the placement of dedicated neutrino detectors in the far-forward region to enable neutrino measurements in a new kinematic regime, in which neutrinos with energies at and beyond 1 TeV interact, thus extending the conventional accelerator-based results. This region is presently not covered by the acceptance of the existing central LHC detectors, namely ATLAS~\cite{Aad:2008zzm}, CMS~\cite{Chatrchyan:2008zzk}, ALICE~\cite{Aamodt:2008zz}, and LHCb~\cite{Alves:2008zz}.

In 1984, future hadron colliders were pointed out as potentially interesting sources of an intense forward high energy neutrino flux dominated by prompt charm and bottom hadron decays~\cite{DeRujula:1984pg}, followed by experimentally interesting order-of-magnitude flux and interaction rate estimates for $\nu_e$, $\nu_\mu$, and in particular $\nu_\tau$ interactions based on simple downstream detector geometries~\cite{DeRujula:1992sn}. The study was later revisited for the LHC, with event rate estimates given for detector locations and geometries relevant to the LHC infrastructure~\cite{Park:2011gh}. The exploration of these ideas for an experimental neutrino programme at the LHC started with studies using various locations for potential detectors~\cite{Beni:2019gxv}, by the XSEN collaboration~\cite{XSEN:2019bel} and with a pilot project set up by the FASER collaboration, which reported the first candidate neutrino interactions~\cite{FASER:2021mtu} in data taken in 2018 during the last year of LHC Run~2. Both XSEN and the FASER project considered emulsion detector technology for recording neutrino interactions, in particular because of the good $\nu_\tau$ identification potential of such a detector.

The LHC Run~3 started in 2022, at a time by which the landscape had considerably developed. By then, two new experiments in the forward region, FASER and SND@LHC, had been proposed, approved, and installed, and were ready to take data when $pp$ collisions resumed at the LHC. Both experiments very quickly reported the first direct observation of neutrino interactions in $pp$ collisions at $\sqrt{s}=13.6~\mathrm{TeV}$~\cite{FASER:2023zcr,SNDLHC:2023pun}. These observations marked the beginning of a new era in neutrino physics, ``The Dawn of Collider Neutrino Physics''~\cite{Worcester:2023njy}.

Beyond the first observations, Run~3 has already enabled the first quantitative measurements with collider neutrinos, including the observation of different neutrino flavours as well as the first energy spectrum and cross section measurements using accelerator-produced neutrinos in the TeV energy range. The current experiments demonstrate, above all, the feasibility of studying neutrino physics at the LHC, and they are also paving the way for the next phase of this programme, which will make use of the High Luminosity LHC (HL-LHC), expected to start operation around 2030.

\begin{figure}
\centering
\includegraphics[width=0.80\textwidth]{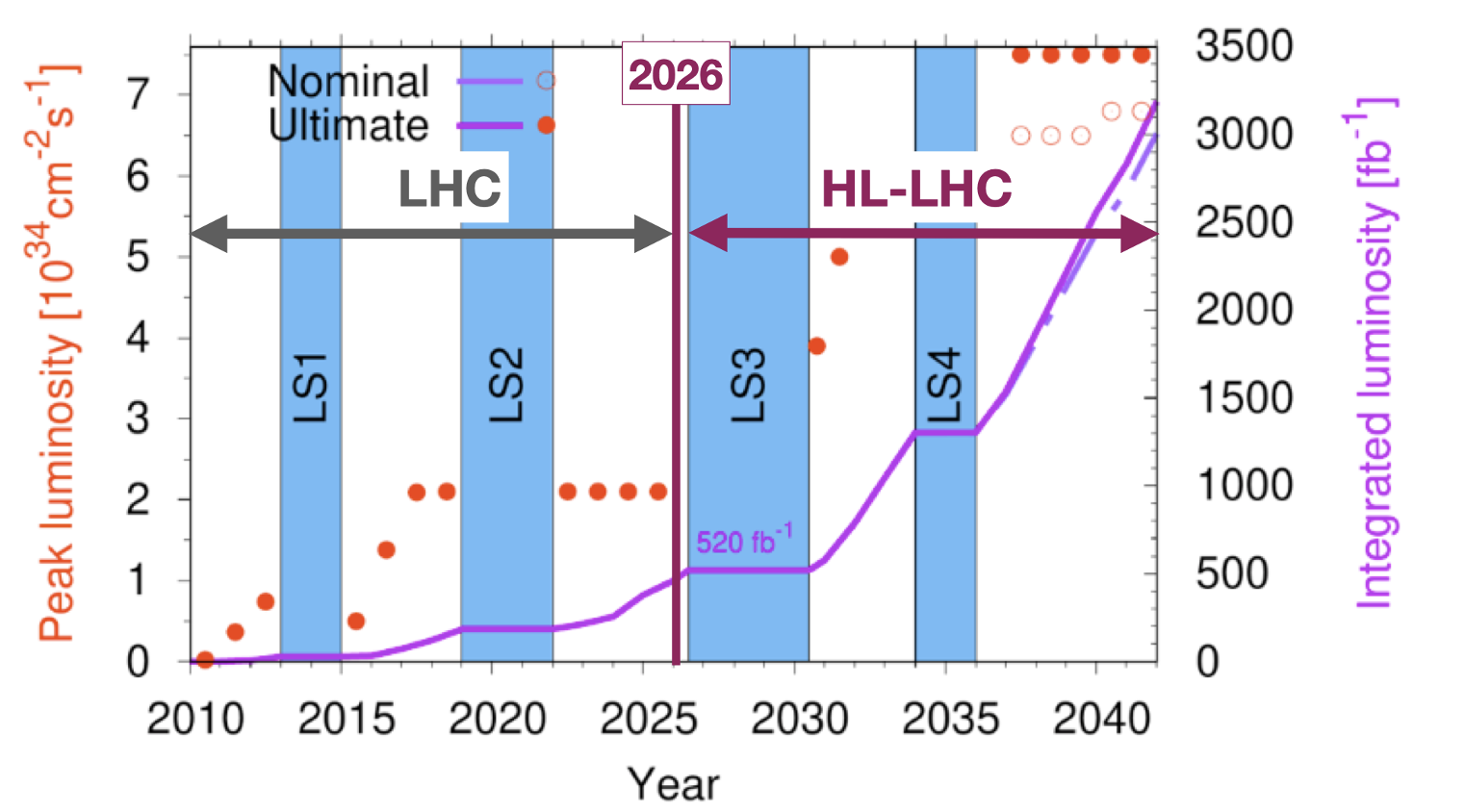}
\caption{Expected LHC luminosity profile, updated by the authors to reflect the LHC schedule as of April 2026. Adapted from Ref.~\cite{BejarAlonso:2020hup} (CC BY 4.0).}
\label{fig:LHC}
\end{figure}

According to the current schedule, the HL-LHC is expected to deliver an integrated luminosity of ${\cal O}(2.5\text{--}3)~\mathrm{ab}^{-1}$, providing the dataset needed to move from first observations to precise differential cross section measurements. The expected luminosity profile of the LHC for the coming years is shown in Fig.~\ref{fig:LHC}. In parallel, upgrades of the current forward experiments have been approved, and several far-forward experimental options are being studied, including a dedicated facility and larger detector systems, in order to increase acceptance and expand the scope of the forward physics programme.

This review article summarises the status of the LHC Run~3 results obtained by the experiments so far and outlines the path towards the HL-LHC. It discusses both the reach of the proposed detector upgrades and additional opportunities such as the Forward Physics Facility and other alternative approaches to neutrino studies at the LHC. It provides an update of the review presented in \cite{Ariga:2025qup}. We also discuss the broader physics opportunities offered by the neutrino study programme.

\section{Forward neutrinos at the LHC}
\label{sec:lhc_neutrino_beam}

\begin{figure}
\centering
        \includegraphics[width=0.90\textwidth]{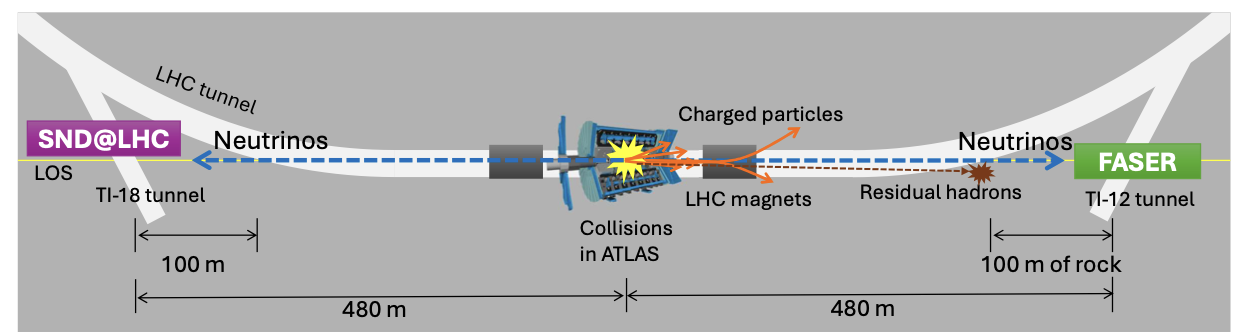}
\caption{Layout of the forward experiments around the ATLAS detector at the LHC. Adapted from Ref.~\cite{Ariga:2025qup} (CC BY 4.0).}
\label{fig:LHC_sketch}
\end{figure}

Neutrinos are produced in essentially every $pp$ collision at the LHC. In the forward direction, the flux is strongly collimated around the collision axis, which requires detectors to be placed along the beam line of sight at a suitable distance from the interaction point. Due to the available infrastructure and space in service tunnels, the forward region downstream of the ATLAS interaction point was chosen as the location for both the FASER and SND@LHC experiments, as shown in Fig.~\ref{fig:LHC_sketch}.

\subsection{Geometry and acceptance}
\begin{figure}
\centering
        \includegraphics[width=0.80\textwidth]{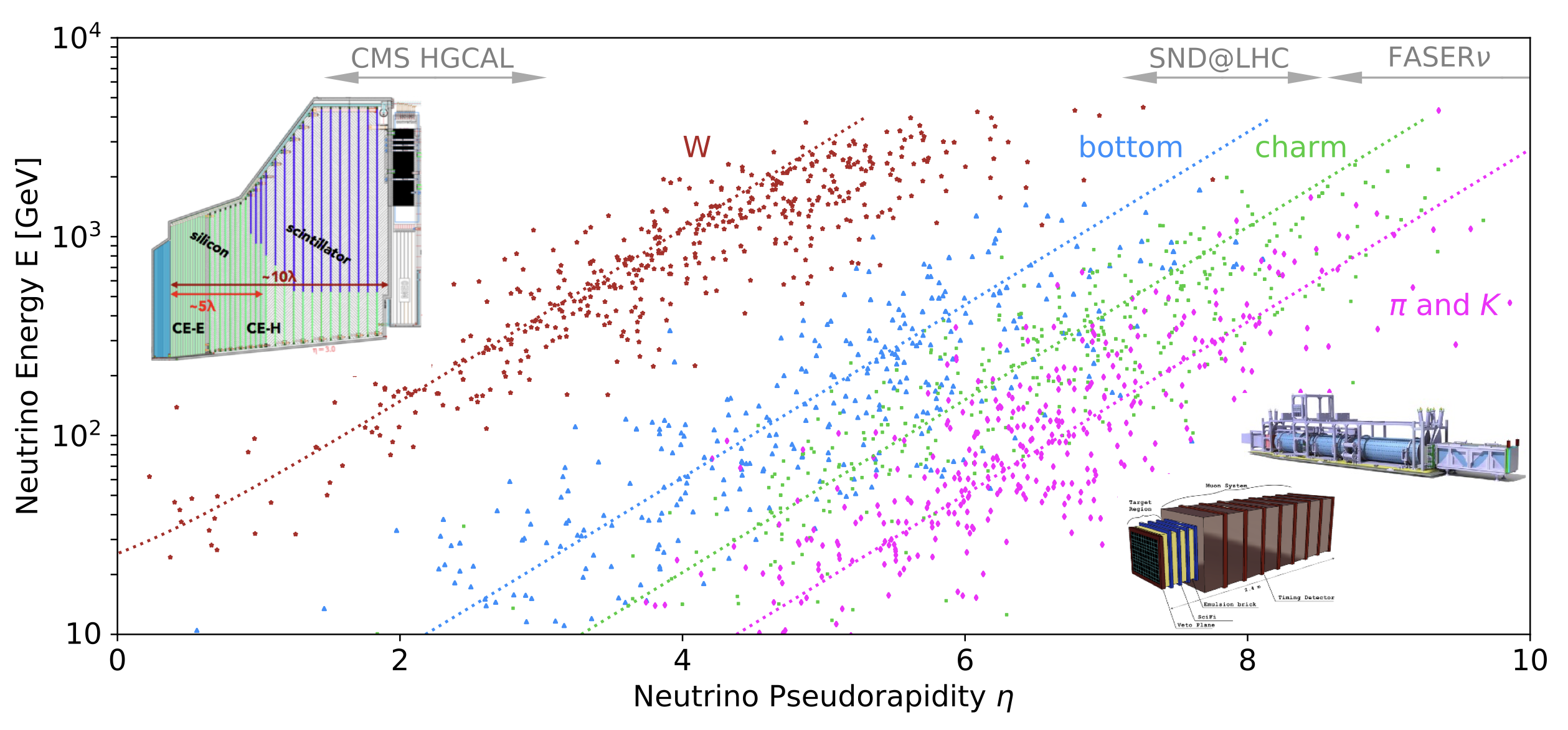}
\caption{Neutrino distribution in the $(\eta,E)$ plane, simulated with the PYTHIA~8 program and weighted by the interaction probability. Colors denote neutrinos from $W$ decays, bottom hadrons, charm hadrons, and light hadrons. Adapted from Ref.~\cite{Foldenauer:2021gkm} (CC BY 4.0); insets follow the original.}
\label{fig:nu_etavsE}
\end{figure}

The forward region is usually described in terms of the variable pseudorapidity, $\eta=-\ln\tan(\theta/2)$, where $\theta$ is the polar angle with respect to the beam line. For $\eta \geq 7$, the corresponding angles are at the mrad level. For small $\theta$, one has $\theta \simeq 2e^{-\eta}$, so that $\eta=7$ corresponds to $\theta \sim 2\times10^{-3}$ rad. For detectors located hundreds of metres downstream of the ATLAS interaction point, such small angles translate into small transverse displacements, $r \simeq L\theta$. At $L \sim 480$ m, a 1 mrad angle corresponds to $r \sim 0.5$ m, which is why compact transverse detectors in existing service tunnels can intercept a sizable fraction of the far-forward flux. Since the neutrino energies are typically in the range of a few hundred GeV up to a few TeV, and therefore the neutrino--nucleus interactions are relatively large, a total target mass of a few tons of material is sufficient to collect a significant number of neutrino interactions. The FASER and SND@LHC design studies discuss how the accessible $\eta$ range is set by the distance from the interaction point together with the available transverse aperture, and how on-axis and slightly off-axis configurations lead to different acceptances and neutrino spectra and benefits~\cite{FASER:2018bac,FASER:2022hcn,SHiP:2020sos,SNDLHC:2022ihg}. The resulting correlation between neutrino energy and pseudorapidity, and the separation of production channels in the $(\eta,E)$ plane, is illustrated in Fig.~\ref{fig:nu_etavsE}.

\subsection{Production mechanisms}
The forward neutrino flux arises from several production channels. Light mesons are copiously produced in $pp$ collisions at LHC energies, leading to some $10^{17}$--$10^{18}$ charged pions per year. Charged pions, charged kaons, and long-lived neutral kaons have mean lifetimes of 12--50~ns, while short-lived neutral kaons decay in less than a nanosecond in their rest frame. At LHC energies, the large relativistic boost increases the decay lengths of these particles in the laboratory frame. Over a distance of a few hundred metres, only a small fraction of these long-lived mesons decay before being absorbed or deflected, and populate a substantial part of the neutrino energy spectrum reaching the forward detectors. In practice, however, this contribution depends strongly on the details of the LHC beam line. Charged pions and kaons are quickly bent away by the accelerator optics and are likely to interact before decaying, while neutral hadrons can continue along the beam direction. Neutrinos and antineutrinos from light hadron decays therefore become most relevant at lower neutrino energies, and their yield is very sensitive to the detector location relative to the first magnetic machine elements and absorbers~\cite{Kling:2021gos}.

A detailed beam simulation, provided by the CERN Sources, Targets and Interactions Group and based on the FLUKA toolkit~\cite{Ballarini:2024isa}, is used to propagate the particles produced in $pp$ interactions at the ATLAS interaction point towards the forward neutrino detectors. The simulation includes all magnetic elements as well as the material of and surrounding the beam pipe. For the purpose of providing a faster turnaround in the simulation chain, a useful fast parametrized simulation was developed in Ref.~\cite{Kling:2021gos}.

The most significant backgrounds for many studies stem from muons originating from meson decays in the forward region, either from particles produced at the $pp$ interaction point or from hadronic/muon re-interactions in accelerator components. Such muons can be a direct background to, for example, muon neutrino measurements, or can produce neutral particles through DIS scattering on nuclei near the detector, which may then enter and interact or decay in the target. Such putative backgrounds are estimated using the detailed beamline simulation program mentioned above.

In the far-forward region relevant for current collider neutrino measurements, prompt decays of heavy flavour hadrons produced at the interaction point, in particular charm and bottom hadrons, provide an important contribution to the neutrino flux, especially at high energies. Because these hadrons decay promptly, their neutrino yield is largely insensitive to downstream magnetic fields and absorbers, and is primarily set by heavy hadron production at very forward rapidities~\cite{Kling:2021gos}.

Electroweak processes contribute as well. Neutrinos from $W$ and $Z$ production mainly populate the highest energy tail and become increasingly important at the larger neutrino angles. The production of $\nu_\tau$ is closely tied to forward produced charm, with the dominant prompt contribution coming from $D_s^\pm$ meson production followed by the leptonic decay $D_s^\pm \to \tau^\pm \nu_\tau\,(\bar{\nu}_\tau)$. The subsequent $\tau$ decay produces an additional $\nu_\tau$ or $\bar{\nu}_\tau$, so that the $\nu_\tau$ yield directly tracks the modelling of forward $D_s$ production and charm hadronisation~\cite{FASER:2018bac,FASER:2019dxq}.

\subsection{Uncertainties and modelling}

The main limitation in predicting forward neutrinos at the LHC is the modelling of forward hadron production, for both light and heavy hadrons, at low transverse momentum and very large rapidity~\cite{Kling:2021gos,FASER:2024ykc}. Different physics assumptions among event generators, choices in hadronisation models, parton distribution function (PDF) uncertainties at small $x$, and QCD scale uncertainties can lead to a sizeable spread in the predicted neutrino fluxes and their flavour dependence.

Phenomenological models are used for light hadron production in predictions for the FASER and SND@LHC experiments, including EPOS-LHC~\cite{Pierog:2013ria}, SIBYLL~2.3d~\cite{Riehn:2019jet}, QGSJET-II-04~\cite{Ostapchenko:2010vb}, DPMJET-III~\cite{Roesler:2000he}, and the PYTHIA~8.3 model tuned to forward hadron production, PYTHIAforward~\cite{Bierlich:2022pfr,Fieg:2023kld}. For neutrinos originating from light hadron decays, predictions from the models considered by FASER agree within roughly 10\% for integrated event rates in the FASER acceptance as evaluated in a dedicated study~\cite{FASER:2024ykc}. A recent SND@LHC analysis comparing DPMJET-III, EPOS-LHC and SIBYLL~2.3d predictions reached a similar conclusion~\cite{SNDLHC:2026oxu}. Measurements from forward experiments such as LHCf~\cite{Piparo:2023yam,LHCf:2020hjf}, which detects very forward neutral particle production, provide important constraints on light hadron production in this region, and the phenomenological models mentioned above are continuously updated as new data become available.

Forward charm production is commonly modelled using perturbative-QCD calculations. In the FASER rate predictions, POWHEG matched to PYTHIA~8.3 is used, with the dominant associated uncertainty estimated from variations of the QCD factorisation and renormalisation scales~\cite{Alioli:2010xd}, which can be as large as 50\% in some regions of phase space. The resulting flux uncertainty is strongly flavour dependent and is particularly important for $\nu_\tau$, whose LHC flux originates predominantly from charm hadron decays. The NA65/DsTau experiment at the CERN SPS is currently analysing its recorded data and aims to make a precise measurement of $D_s$ production in proton--nucleus collisions, which will provide an important benchmark for forward charm and neutrino flux modelling~\cite{DsTau:2025}.

A second source of uncertainty is the modelling of neutrino interactions in the detector target. The LHC neutrino experiments use heavy nuclear targets, in particular tungsten, so the interactions occur on nuclei. Experimental cross sections are often reported per nucleon by normalising to the number of target nucleons. Since this convention has been widely adopted by the LHC experiments so far, we retain the neutrino--nucleon or per-nucleon terminology throughout this review when referring to the quantities reported by the experiments.

At the neutrino energies relevant for the LHC, deep inelastic scattering (DIS) dominates the inclusive interaction cross section, with uncertainties arising primarily from PDFs and, for heavy nuclear targets, from nuclear effects~\cite{CooperSarkar:2011pa}. Comparisons of different cross section calculations used for FASER indicate uncertainties of approximately 6\% for neutrino energies above 100~GeV~\cite{FASER:2024ykc}. A theory effort is ongoing to combine the well developed elements from the different calculations in order to obtain the most complete high energy neutrino DIS interaction calculation.

As one example, used later in comparisons with the experimental results, the Bodek--Yang model provides a phenomenological description of inclusive neutrino--nucleon scattering based on effective leading order PDFs~\cite{Bodek:2002vp,Bodek:2004pc,Bodek:2010km}. It introduces a modified scaling variable, $\xi_w$, together with phenomenological $K$ factors at low $Q^2$ to account for non-perturbative effects and to extend the description from the DIS regime towards lower $Q^2$ and the resonance region. In measurements that extract neutrino cross sections or fluxes, these interaction model uncertainties enter together with detector effects such as acceptance, efficiency, and energy response. Several of the cross section predictions shown in this review are based on the Bodek--Yang model, as implemented in the GENIE neutrino interaction Monte Carlo generator~\cite{Andreopoulos:2009rq,Andreopoulos:2015wxa}.

Given all these uncertainties, together with ongoing efforts to use forward data to improve the modelling of forward hadron and neutrino production, the use of different detector geometries and technologies in the far-forward region is important and complementary. On-axis and off-axis configurations probe different parent hadron mixtures and provide complementary handles on backgrounds and interaction topologies, as discussed in the next section.

\section{Experiments in Run~3}
\label{sec:experiments}
During the LHC Run~3, the collider neutrino programme used dedicated detectors installed in the existing service tunnels downstream of the ATLAS interaction point. FASER sits on the collision axis in the TI12 tunnel, while SND@LHC operates off-axis in TI18, on the opposite side of ATLAS. At these locations the LHC magnetic lattice sweeps away charged particles and the shielding and rock around the accelerator absorb most hadrons, while neutrinos continue to propagate essentially unaffected to both experimental sites. The residual background is dominated by penetrating muons produced at the interaction point or in upstream interactions and meson decays, together with secondary neutral particles generated by muon interactions in the surrounding material on the way to the detector. The experimental locations, on-axis along the line of sight in TI12 for FASER and off-axis in TI18 for SND@LHC, lead to different acceptances and background conditions, making the measurements of the two experiments complementary.

Both detectors described in more detail below are hybrid detectors. They consist of active electronic detectors, which record and time-stamp individual interactions occurring in the setup, and passive emulsion detectors, in which emulsion bricks integrate particle tracks over a given run period, typically 1--2 months in Run~3. After exposure, the emulsion films need to be chemically developed, scanned with automatic microscope systems, and reconstructed in three dimensions. The emulsion stacks are typically replaced three to four times per year, but their processing and analysis are rather laborious and time consuming. Therefore, the first results from these experiments were obtained using the electronic detectors only.

\subsection{FASER experiment}
\begin{figure}
\centering
\includegraphics[width=0.60\textwidth]{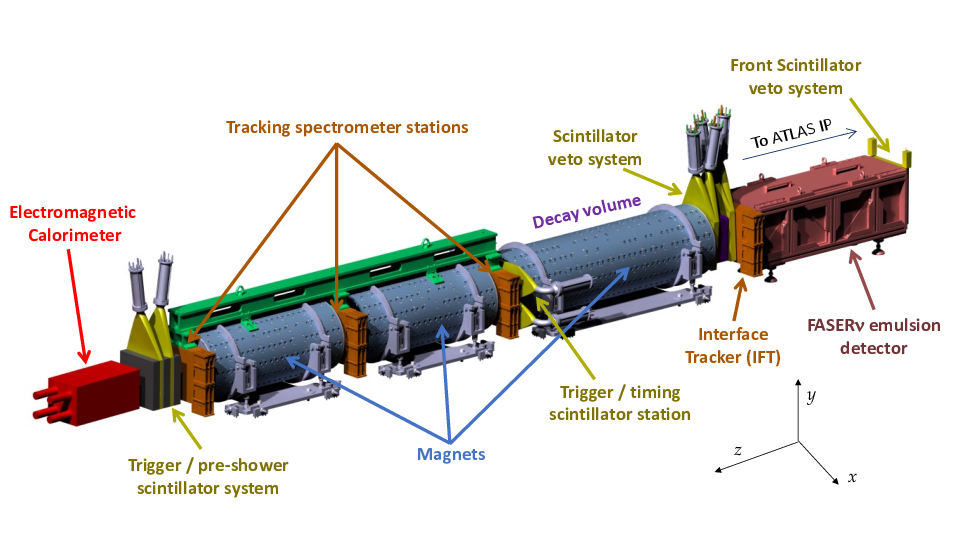}
\caption{Schematic layout of the FASER detector, including the upstream FASER$\nu$ emulsion target, the decay volume for dark sector particle searches, the magnetic spectrometer with tracking stations, and the electromagnetic calorimeter. Adapted from Ref.~\cite{FASER:2022hcn} (CC BY 4.0).}
\label{fig:faser_schematic}
\end{figure}

FASER (ForwArd Search ExpeRiment) is a compact magnetic spectrometer installed on the collision axis and designed to reconstruct forward charged particles with good vertexing and momentum resolution~\cite{FASER:2018bac,FASER:2022hcn}. The idea of searching for light, weakly coupled new particles in the very forward region of LHC collisions was proposed in Ref.~\cite{Feng:2017uoz} in 2017. After a collaboration had been formed, the baseline new particle search spectrometer, FASER, was approved in March 2019. In October 2019, the experiment was supplemented with an emulsion-based target for detecting neutrino interactions, FASER$\nu$, as described in Ref.~\cite{Boyd:2026xuf}. The detector is located 480 m from the ATLAS interaction point and uses scintillator stations for triggering and vetoing incoming charged tracks, tracking stations in a permanent dipole 0.6~T magnetic field to measure momenta and directions, and an electromagnetic calorimeter for energy measurements and particle identification. This apparatus serves a dual role. FASER was originally designed to be sensitive to new light, weakly interacting long-lived particles that can be produced in meson decays, but it also acts as an excellent forward spectrometer for neutrino interactions occurring in a dedicated upstream target~\cite{FASER:2019dxq}. A schematic layout of the FASER detector and the location of the FASER$\nu$ target are shown in Fig.~\ref{fig:faser_schematic}.

The neutrino programme uses the dedicated FASER$\nu$ target placed directly in front of the spectrometer. FASER$\nu$ is based on nuclear emulsion films interleaved with tungsten plates, providing a dense target together with micron-scale tracking and vertex reconstruction. The emulsion detector integrates particle tracks over an exposure period and is then removed for offline scanning and reconstruction~\cite{FASER:2025reco}. Neutrino candidates are identified through a reconstructed primary vertex inside the target, together with outgoing charged tracks. The emulsion detector also enables momentum measurements of charged particles produced in neutrino interactions using multiple Coulomb scattering~\cite{FASER:2026mcs}. For $\nu_\mu$ charged-current (CC) events, an outgoing muon can be matched to a penetrating track reconstructed in the downstream spectrometer, providing strong background rejection and a clean topology for the first observations. The electronic spectrometer data further enable complementary measurements based on event-by-event information, including energy dependent studies with future larger statistics datasets.

\subsection{SND@LHC experiment}
SND@LHC (Scattering and Neutrino Detector at the LHC) originated from the merger of ideas developed in the XSEN proposal~\cite{XSEN:2019bel} and in the proposed SND detector for neutrino detection in the SHiP fixed target beam-dump experiment at CERN~\cite{SHiP:2021nfo}. At the time, the future of SHiP was still uncertain. Therefore, in 2020 it was decided to propose a dedicated neutrino study programme at the LHC, based on the studies carried out for these two projects. The emerging experiment, SND@LHC, was approved in 2021 and installed in 2022, in time for the start of Run~3 at the LHC. Because of this ``just-in-time'' installation, there was little opportunity to modify the tunnel environment at the detector location. For example, a different layout could in principle have allowed a larger acceptance or a more complete background veto configuration. However, such modifications were not feasible on the available timescale. As a result, the detector design represents a compromise between rapid installation and physics reach, with some limitations on the number of neutrino events that could be collected during the first years of operation. Since SHiP was approved in March 2024, SND@LHC also serves as a useful pilot project for the future SHiP neutrino programme.

\begin{figure}
\centering
\includegraphics[width=0.60\textwidth]{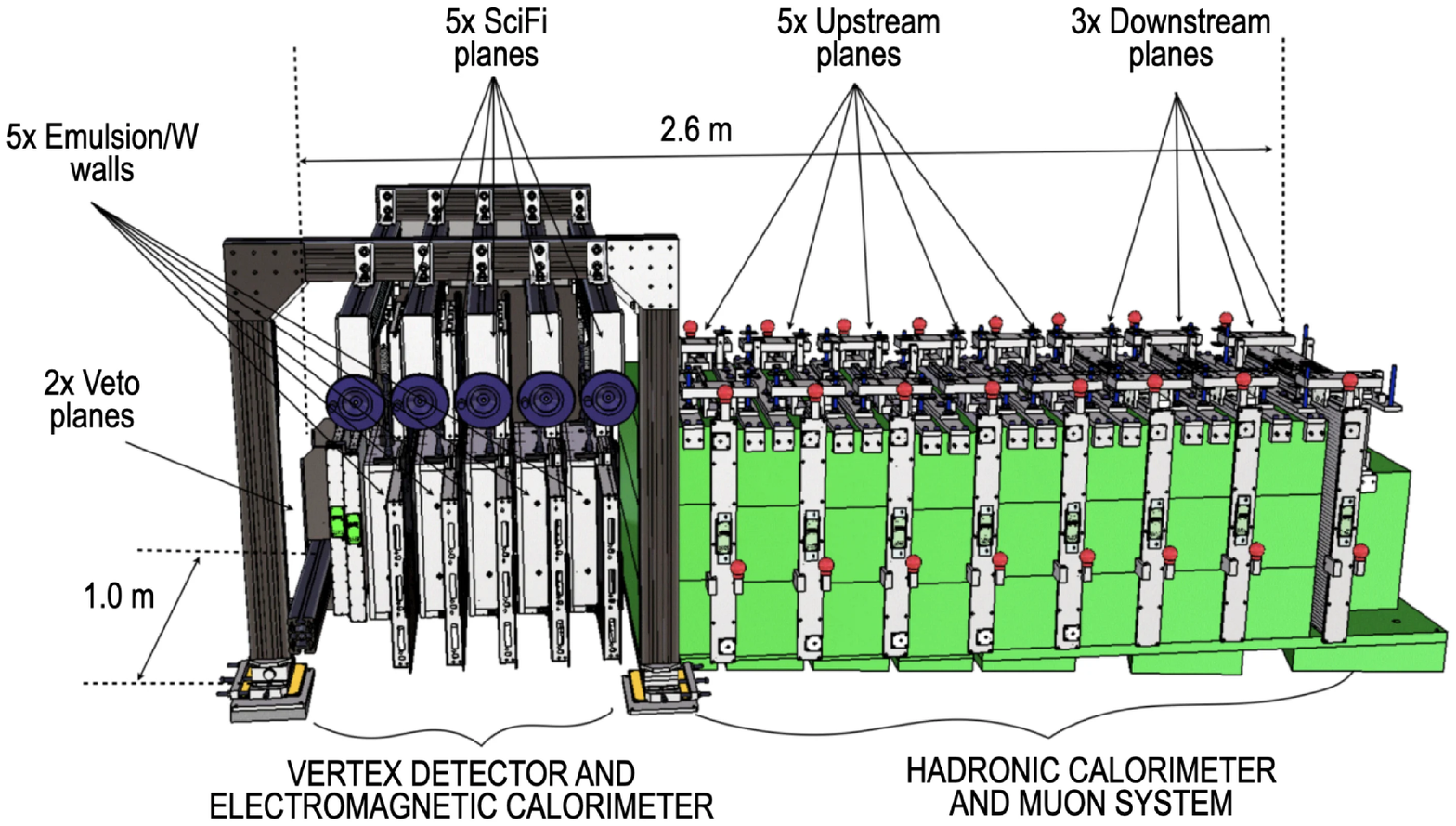}
\caption{Schematic layout of the SND@LHC detector, including the upstream veto system, the emulsion--tungsten target and vertex detector with scintillating fibre (SciFi) tracking planes, and the downstream hadronic calorimeter and muon identification system. Adapted from Ref.~\cite{SNDLHC:2022ihg} (CC BY 4.0).}
\label{fig:snd_schematic}
\end{figure}

The SND@LHC experiment is installed in TI18, approximately the same longitudinal distance from the ATLAS interaction point as FASER, about 480~m, but with an off-axis acceptance. Its design combines a dense target region with emulsion and electronic readout. The emulsion target, consisting of nuclear emulsion films interleaved with tungsten (W) plates -- so-called Emulsion Cloud Chambers (ECCs) -- provides precise vertex reconstruction, while the electronic tracking planes based on scintillating fibre (SciFi) provide timestamps that allow event association in the high-rate far-forward environment~\cite{SHiP:2020sos,SNDLHC:2022ihg}. Downstream of the target, SND@LHC includes a hadronic calorimeter and a muon identification detector based on scintillator planes embedded in an iron absorber structure. The electronic tracking and scintillator systems provide time information that is used for event association and background rejection. A schematic view of the SND@LHC detector layout and its main subsystems is shown in Fig.~\ref{fig:snd_schematic}.

\section{Run~3 results}
\label{sec:run3_results}
Run~3 has established collider neutrinos as an experimentally accessible signal for the first time and has already produced the first quantitative measurements in the far-forward region at the LHC. The initial analyses naturally targeted the cleanest event topologies, where a CC interaction produces a penetrating muon that can be tagged with high purity and efficiency and used to suppress the dominant muon-induced backgrounds. With these observation channels in place, the programme has moved to first measurements of CC cross sections in the TeV energy range. More recently, Run~3 has also delivered the first neutrino interaction samples without final state muons, opening access to NC and $\nu_e$-enriched event topologies and providing a complementary constraint on the forward neutrino flux. Promising results from emulsion data, although still based on modest fractions of the total recorded sample, have also been released by the FASER collaboration.

\subsection{First observations of collider neutrinos}
\label{subsec:first_obs}
For these first observations, both experiments started from the most robust signature, a search for $\nu_\mu$ CC interactions tagged by a penetrating muon, which strongly suppresses the dominant muon-induced backgrounds. FASER and SND@LHC implemented this idea in different geometries and with different detector technologies, leading to complementary acceptances and background compositions.

\begin{figure}
\centering
\includegraphics[width=0.54\textwidth]{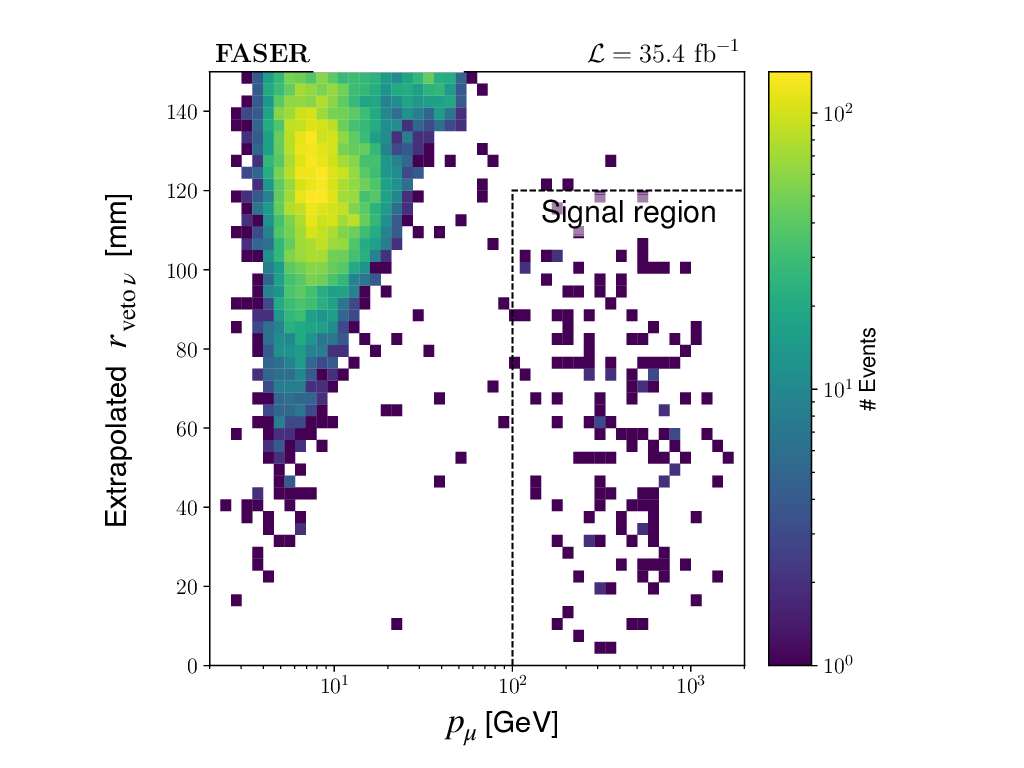}
\includegraphics[width=0.40\textwidth]{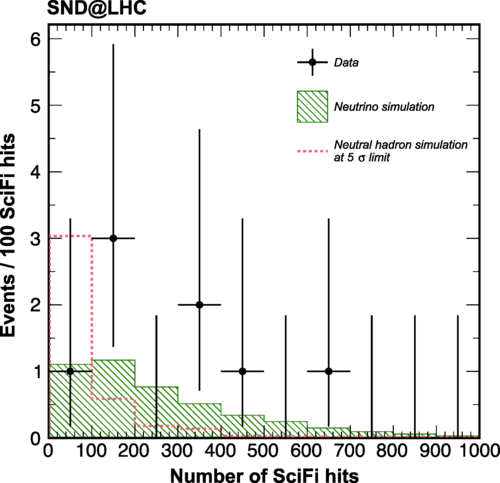}
\caption{Left: distribution of the extrapolated track position at the FASER$\nu$ scintillator plane, $r_{\mathrm{veto}\,\nu}$, versus the reconstructed muon momentum $p_\mu$ for the 2022 dataset. The dashed lines indicate the signal region definition used in the FASER $\nu_\mu$ CC analysis. Adapted from Refs.~\cite{FASER:2023zcr} (CC BY 4.0). Right: distribution of the number of SciFi hits for the selected sample in the SND@LHC $\nu_\mu$ CC analysis, comparing data to the neutrino simulation and to the neutral hadron background estimate. Adapted from Ref.~\cite{SNDLHC:2023pun} (CC BY 4.0).}
\label{fig:firstobs_panels}
\end{figure}

\begin{itemize}
  \item \textbf{FASER:} Using 2022 $pp$ data at $\sqrt{s}=13.6~\mathrm{TeV}$ corresponding to $35.4~\mathrm{fb}^{-1}$, FASER identified $\nu_\mu$ CC candidates with a penetrating muon reconstructed in the magnetic spectrometer and consistent with an interaction in the upstream target region. A key feature of the FASER strategy was the use of an extrapolated track position at the FASER$\nu$ scintillator plane to define a signal-enhanced region close to the ATLAS line of sight, combined with a momentum requirement to suppress geometric backgrounds from charged particles that missed the upstream veto. Backgrounds were constrained with sidebands and control samples, and the signal yield was extracted with a likelihood fit, yielding $153^{+12}_{-13}$ inferred neutrino interactions over the background expectation, with a significance of $16\sigma$~\cite{FASER:2023zcr}. The signal region definition is illustrated in Fig.~\ref{fig:firstobs_panels} (left).

  \item \textbf{SND@LHC:} Using 2022 $pp$ data at $\sqrt{s}=13.6~\mathrm{TeV}$ corresponding to $36.8~\mathrm{fb}^{-1}$, SND@LHC identified $\nu_\mu$ CC candidates with an interaction in the target region and a penetrating muon track traversing the full muon system, reconstructed with the active electronic components. A key feature of the SND@LHC strategy was the use of event topology and activity in the SciFi tracker together with veto information to suppress incoming charged particle backgrounds and muon-induced secondaries. After selection, 8 candidates remained with an estimated background of 0.086 events, yielding a significance of about $7\sigma$~\cite{SNDLHC:2023pun}. The SciFi hit multiplicity used to validate the selection is shown in Fig.~\ref{fig:firstobs_panels} (right).
\end{itemize}

Taken together, the two observations relied on the same muon-tagged $\nu_\mu$ CC signature, but they used different experimental configurations. FASER exploited an on-axis location and a magnetic spectrometer, so that muon pointing and momentum provided a compact signal definition and allowed the residual backgrounds to be constrained with sidebands and a likelihood fit. SND@LHC observed the signal in an off-axis configuration using the active electronic detector, where the selection was driven by event topology in the SciFi tracker and the requirement of a track traversing the full muon system, together with veto-based rejection of incoming activity. Both experimental results are in agreement with the expected number of interactions within the uncertainties. This provides an important first demonstration, and a strong cross check, of the experimental feasibility of performing neutrino physics studies at a collider.

\subsection{First cross section measurements in the TeV regime}
Following the first observations, Run~3 enabled the first measurements of neutrino--nucleon CC cross sections in the TeV energy range using the FASER$\nu$ emulsion--tungsten detector~\cite{FASER:2024hoe}. For this result, a 128.6~kg portion of the FASER$\nu$ target exposed to 9.5~fb$^{-1}$ of 2022 $pp$ collision data at $\sqrt{s}=13.6~\mathrm{TeV}$ was analysed. Candidate $\nu_e$ and $\nu_\mu$ interactions were identified by reconstructing a primary vertex in the emulsion target together with an outgoing charged lepton signature, suppressing the dominant neutral hadron background. With a requirement of $E_e>200~\mathrm{GeV}$, four $\nu_e$ candidates were observed with an expected background of $0.025^{+0.015}_{-0.010}$, corresponding to a significance of 5.2$\sigma$. With a corresponding requirement of $p_\mu>200~\mathrm{GeV}$ for the muon candidate, eight $\nu_\mu$ candidates were observed with an expected background of $0.22^{+0.09}_{-0.07}$, equivalent to a significance of 5.7$\sigma$.

\begin{figure}
\centering
\includegraphics[width=0.45\textwidth]{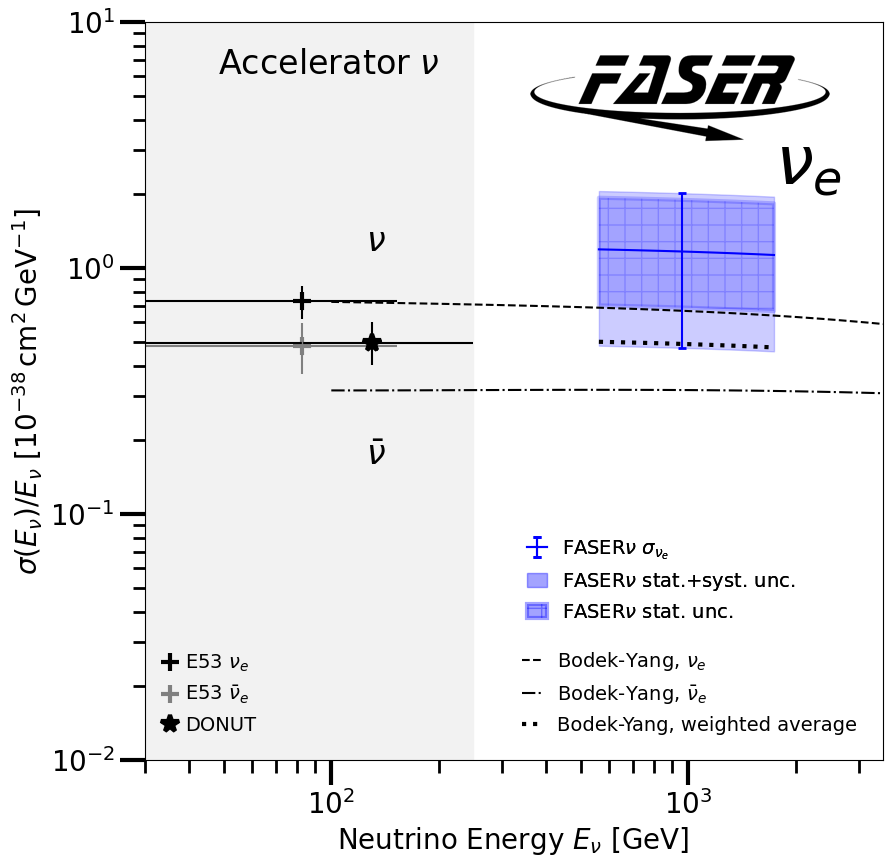}
\includegraphics[width=0.80\textwidth]{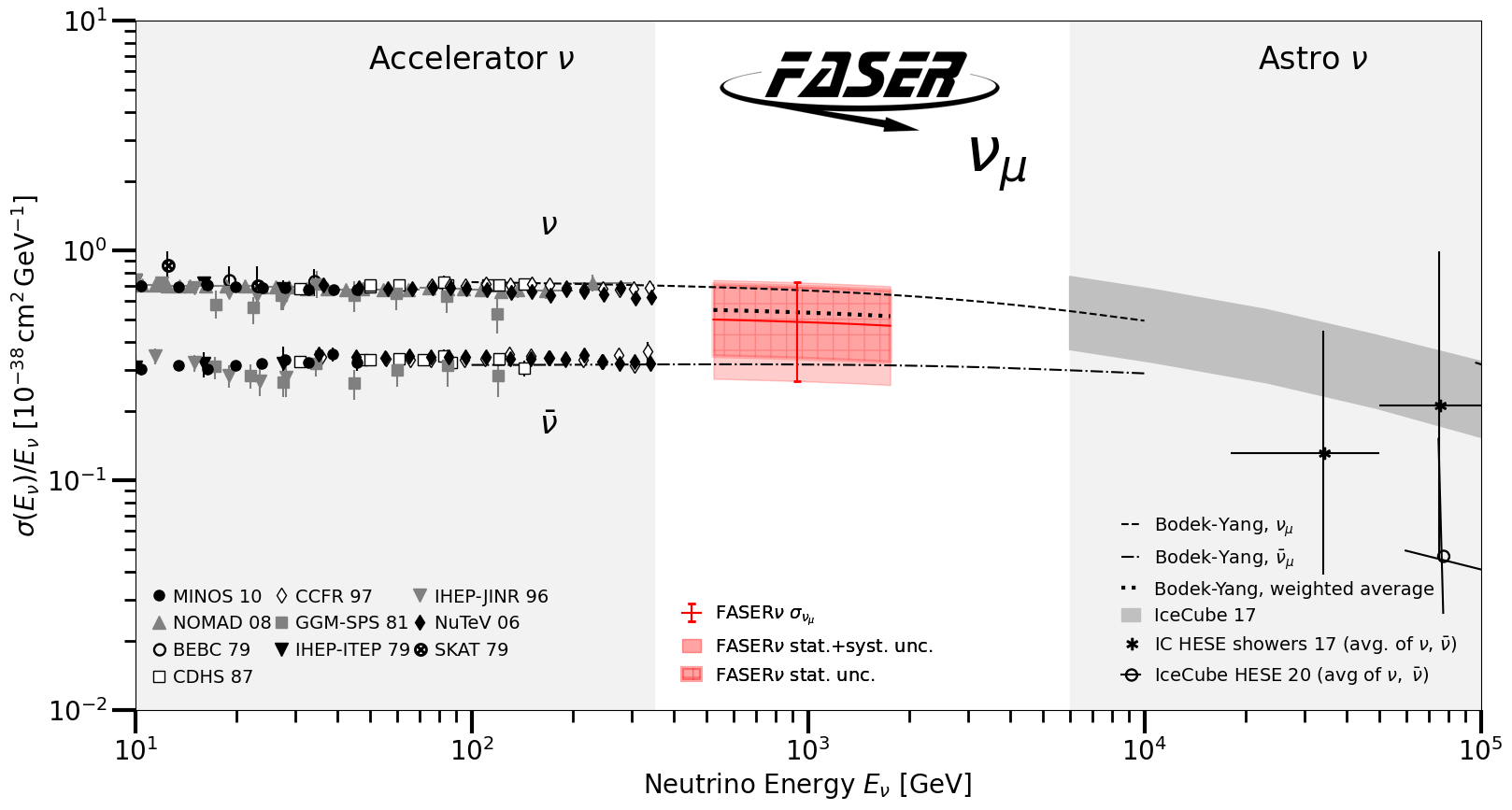}
\caption{Measured $\sigma/E_\nu$ per nucleon for $\nu_e$ (top) and $\nu_\mu$ (bottom) in the TeV energy range, compared to the Standard Model predictions. The dashed contours labeled ``Bodek--Yang'' indicate the Bodek--Yang cross section model used for comparison. Adapted from Ref.~\cite{FASER:2024hoe} (CC BY 4.0).}
\label{fig:faser_xs_tev}
\end{figure}

The measurement combined neutrino and antineutrino interactions, and the results were therefore reported for $(\nu_e+\bar{\nu}_e)$ and $(\nu_\mu+\bar{\nu}_\mu)$. The neutrino--nucleon cross section per energy, $\sigma/E_\nu$, was extracted in a single energy bin, yielding $\sigma/E_\nu = (1.2^{+0.8}_{-0.7})\times 10^{-38}~\mathrm{cm^2\,GeV^{-1}}$ for $\nu_e$ over 560--1740~GeV, and $\sigma/E_\nu = (0.5\pm0.2)\times 10^{-38}~\mathrm{cm^2\,GeV^{-1}}$ for $\nu_\mu$ over 520--1760~GeV, consistent with the Standard Model expectations. The measured cross sections per nucleon are shown in Fig.~\ref{fig:faser_xs_tev}.

\begin{figure}
\centering
\includegraphics[width=0.75\textwidth]{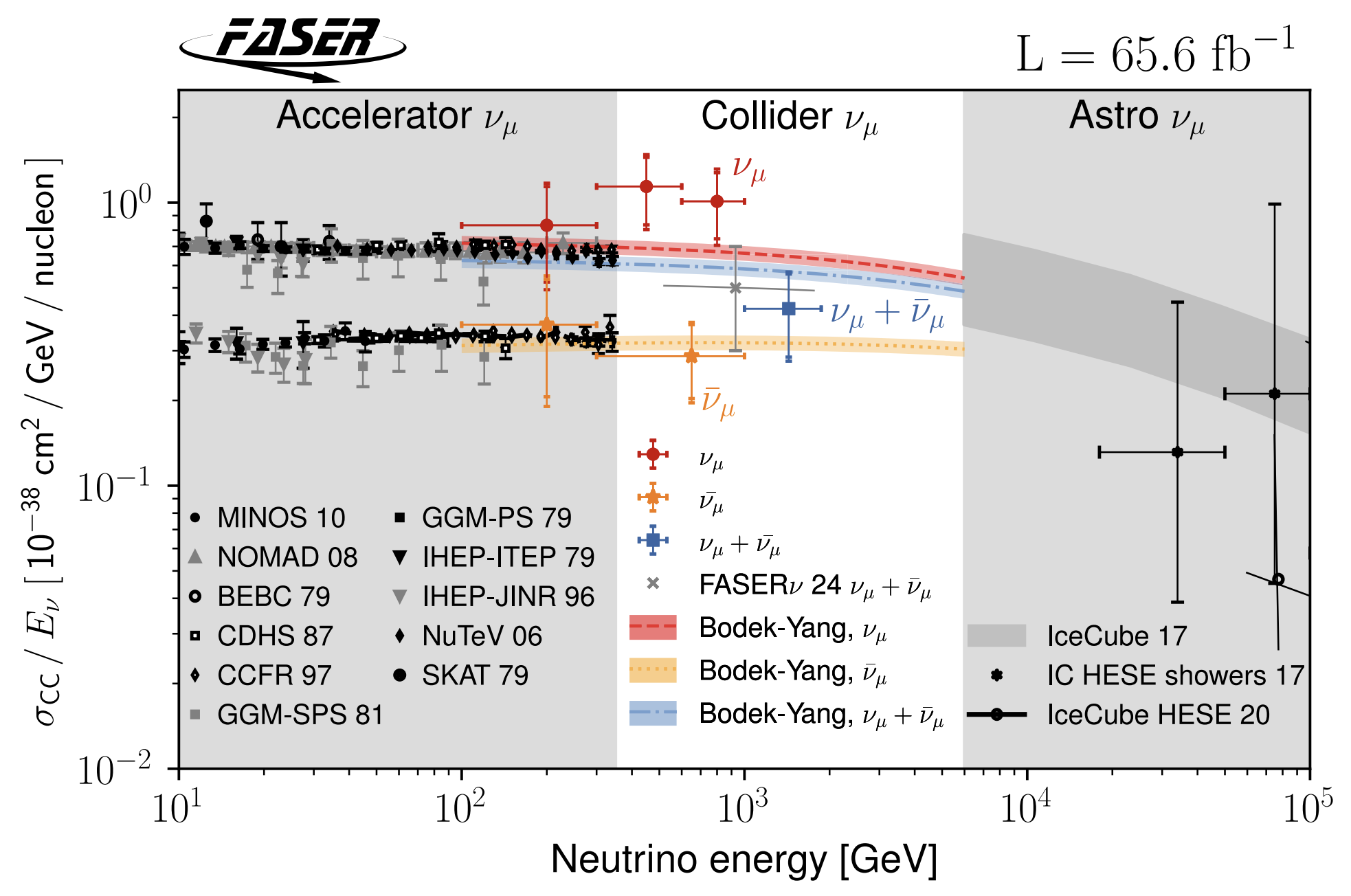}
\caption{Measured $\nu_\mu$ charged-current cross section divided by neutrino energy as a function of neutrino energy, compared to Bodek--Yang predictions and previous neutrino measurements. Adapted from Ref.~\cite{FASER:2024ref} (CC BY 4.0).}
\label{fig:faser_numu_energy}
\end{figure}

Subsequently, FASER extended these measurements using the active electronic detector, identifying $338.1\pm21.0$ $\nu_\mu$ CC interactions in $65.6\pm1.4$~fb$^{-1}$ of Run~3 data. This enabled the first differential measurements of the $\nu_\mu$ interaction cross section and flux as a function of neutrino energy~\cite{FASER:2024ref}. These results provide direct constraints on the energy dependence of the forward neutrino cross section, as shown in Fig.~\ref{fig:faser_numu_energy}, and are complemented by a measurement of the $\nu_\mu$ flux as a function of rapidity~\cite{Conf-FASER-CONF-2025-001}.

\subsection{Neutrino interactions without final state muons}
SND@LHC reported the first observation of neutrino interactions with no explicitly reconstructed final state muon, denoted as $\nu_{0\mu}$ events~\cite{SNDLHC:2024qqb}. The analysis used the 2022--2023 dataset at $\sqrt{s}=13.6~\mathrm{TeV}$, corresponding to an integrated luminosity of 68.6~fb$^{-1}$. It targeted a sample dominated by NC interactions and $\nu_e$ CC interactions, with a small contribution from $\nu_\tau$. 

The event selection proceeded in two steps. Events consistent with a neutral particle interacting in the tungsten target were first selected using fiducial requirements based on the average SciFi hit position, together with a requirement of no hits in the veto system to reject incoming charged particles. Shower-like topologies were then selected by requiring hits in at least two SciFi stations and significant activity in the two most upstream layers of the hadron calorimeter. Reconstructible muons were suppressed by discarding events with hits in the two most downstream planes of the muon detector, ensuring that no muon track can be reconstructed. In the second step, a hit-density weight variable was built from the local density of SciFi hits within 1~cm in a given station. The largest sum of these weights among the SciFi stations was used as the main discriminant. This discriminant was validated with hadron test-beam data. A low hit-density control region was used to constrain the neutral hadron background from muon DIS in the tunnel walls. The signal region threshold was chosen by maximising the expected sensitivity using a one-sided profile likelihood procedure.

\begin{figure}
\centering
\includegraphics[width=0.50\textwidth]{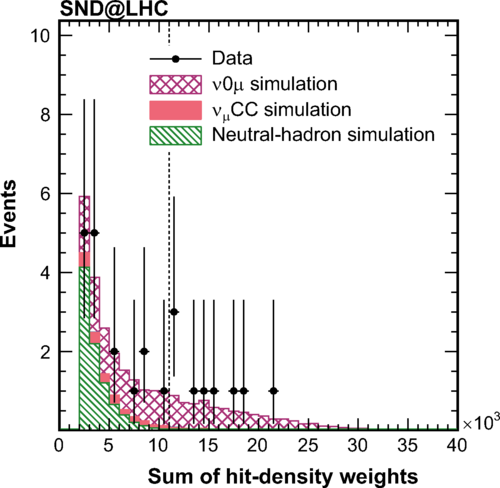}
\caption{Distribution of the sum of SciFi hit-density weights used as the main discriminant in the SND@LHC $\nu_{0\mu}$ analysis, comparing data to the expected signal and background contributions. Adapted from Ref.~\cite{SNDLHC:2024qqb} (CC BY 4.0).}
\label{fig:snd0mu_discriminant}
\end{figure}

After all selection cuts, 9 candidate events were observed with an estimated background of 0.32 events, corresponding to an observation significance of 6.4$\sigma$. The expected $\nu_{0\mu}$ signal was dominated by $\nu_e$ CC and NC interactions, and the analysis also provided evidence for a $\nu_e$ CC component by constraining the NC contribution using the measured $\nu_\mu$ CC rate. The corresponding distribution of the SciFi hit-density discriminant is shown in Fig.~\ref{fig:snd0mu_discriminant}.

\subsection{New results from Run~3}
Several updated and new preliminary results have been reported recently, following the first publications discussed above. They extend the Run~3 collider neutrino programme from first observations towards more differential cross section measurements, and first studies of more exclusive neutrino interaction topologies.

\begin{figure}
\centering
\includegraphics[width=0.70\textwidth]{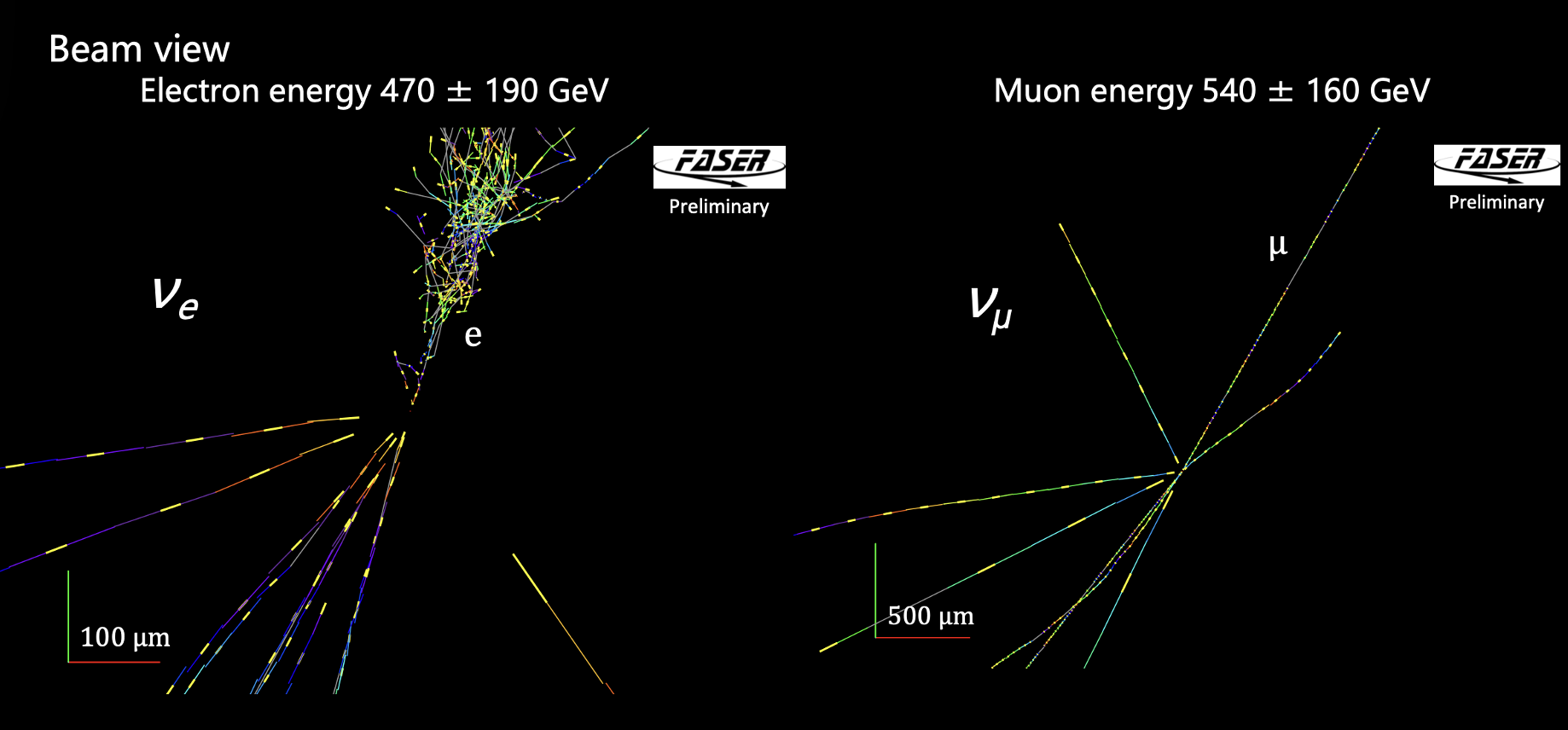}
\caption{A $\nu_e$ and a $\nu_{\mu}$ interaction candidate in the FASER emulsion target. Adapted from Ref.~\cite{Conf-FASER-CONF-2026-002}.}
\label{fig:faser-emulsion}
\end{figure}

\paragraph{FASER.}
FASER has presented updated results on high energy electron and muon neutrino CC interactions in the FASER$\nu$ emulsion--tungsten sub-detector. A subset of the FASER$\nu$ volume was used, corresponding to a target mass of 681.1~kg and an exposure of 9.5~fb$^{-1}$ of LHC $pp$ collision data collected in 2022~\cite{Conf-FASER-CONF-2026-002,Zhang:2026cpk}. The muon neutrino energy is reconstructed using a regression boosted decision tree, and the energy dependent cross section is measured in two bins. The electron neutrino cross section is measured in a single bin; no energy dependent measurement is attempted. The muon and electron neutrino cross section measurements are found to be consistent with earlier published FASER results within experimental uncertainties, both for the analysis using the electronic detector ($\nu_{\mu}$) and for the emulsion detector analyses ($\nu_{\mu},\nu_e$). Clear $\nu_e$ and $\nu_\mu$ interaction candidates in the FASER emulsion target are shown in Fig.~\ref{fig:faser-emulsion}.

\begin{figure}
\centering
\includegraphics[width=1.00\textwidth]{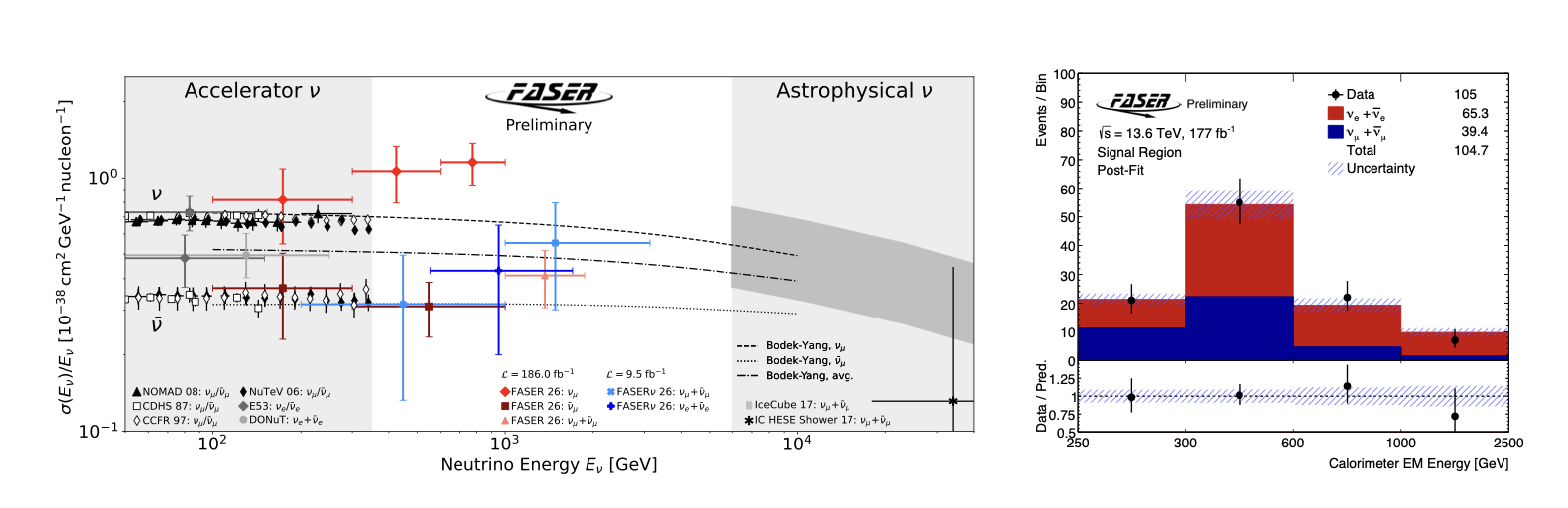}
\caption{Left: measured $\nu_{\mu}$ and $\nu_e$ CC neutrino--nucleon cross sections from the FASER$\nu$ emulsion detector and the FASER electronic detector, compared with theoretical predictions. Right: post-fit calorimeter electromagnetic energy distribution in the $\nu_e$ signal region, showing an excess of $65\pm12$ events above background. Adapted from Ref.~\cite{Zhang:2026cpk}.}
\label{fig:faser-new}
\end{figure}

FASER has also reported preliminary results on the observation of electron neutrinos in the FASER electronic detector~\cite{Conf-FASER-CONF-2026-004}. The analysis uses the electromagnetic calorimeter located downstream at the end of the experiment and is based on the measured electromagnetic energy deposits. It uses the combined 2022, 2023 and 2024 Run~3 $pp$ dataset, corresponding to an integrated luminosity of $(176.8\pm3.6)$~fb$^{-1}$. The analysis considers $\nu_e$ CC and $\nu_e$ NC interactions, together with their antineutrino counterparts, as signal, with a dominant background from muon neutrinos and antineutrinos. Events are selected by requiring large electromagnetic energy deposits in the calorimeter and no activity in the FASER veto scintillator systems. An excess of $65\pm12$ events above the background only expectation is observed, consistent with the expected electron neutrino signal of $42\pm27$ events. The background only hypothesis is therefore rejected with a significance of 5.5 standard deviations. The preliminary FASER cross section and electronic detector $\nu_e$ results are shown in Fig.~\ref{fig:faser-new}.

FASER has further released preliminary results on the first search for charm hadron production in high energy electron and muon neutrino CC interactions with the emulsion--tungsten detector~\cite{Conf-FASER-CONF-2026-003}. Neutrino interactions with charm are an important future physics goal of both FASER and SND@LHC, as discussed in Sec.~\ref{sec:physics}. Dedicated tools for reconstructing displaced secondary vertices and classifying decay topologies of charged charm hadrons have been developed and validated using Monte Carlo simulation. A multivariate analysis using Lorentz invariant kinematic variables is employed to discriminate charm hadron decays from hadronic interaction backgrounds. The analysis is applied to 40 neutrino interaction candidates identified in 9.5~fb$^{-1}$ collected during 2022, and good agreement between data and simulation is observed for the secondary vertex properties, indicating that first physics results on this topic could follow soon.

Finally, FASER has shown that, with the statistics of the 2022--2024 data samples, double-differential measurements of the muon-neutrino flux and the neutrino--nucleon cross section as functions of neutrino energy and rapidity become meaningful~\cite{Conf-FASER-CONF-2026-005}. Charged-current muon neutrino interactions are identified using the active electronic components.

\paragraph{SND@LHC.}
The SND@LHC collaboration has reported an updated measurement of $\nu_{\mu}$ CC interactions using the complete 2022 and 2023 dataset, corresponding to an integrated luminosity of 68.6~fb$^{-1}$~\cite{SNDLHC:2026oxu}. A total of 31 $\nu_{\mu}$ CC candidates are selected, with an expected background of $5.0\pm1.1$ events and a signal expectation of $24^{+10}_{-9}$ events. The combined $\nu_{\mu}$ and $\bar{\nu}_{\mu}$ CC cross section on tungsten is measured to be $(37^{+24}_{-12})\times10^{-35}$~cm$^2$ at a median neutrino energy of 228~GeV.

\begin{figure}
\centering
\includegraphics[width=0.80\textwidth]{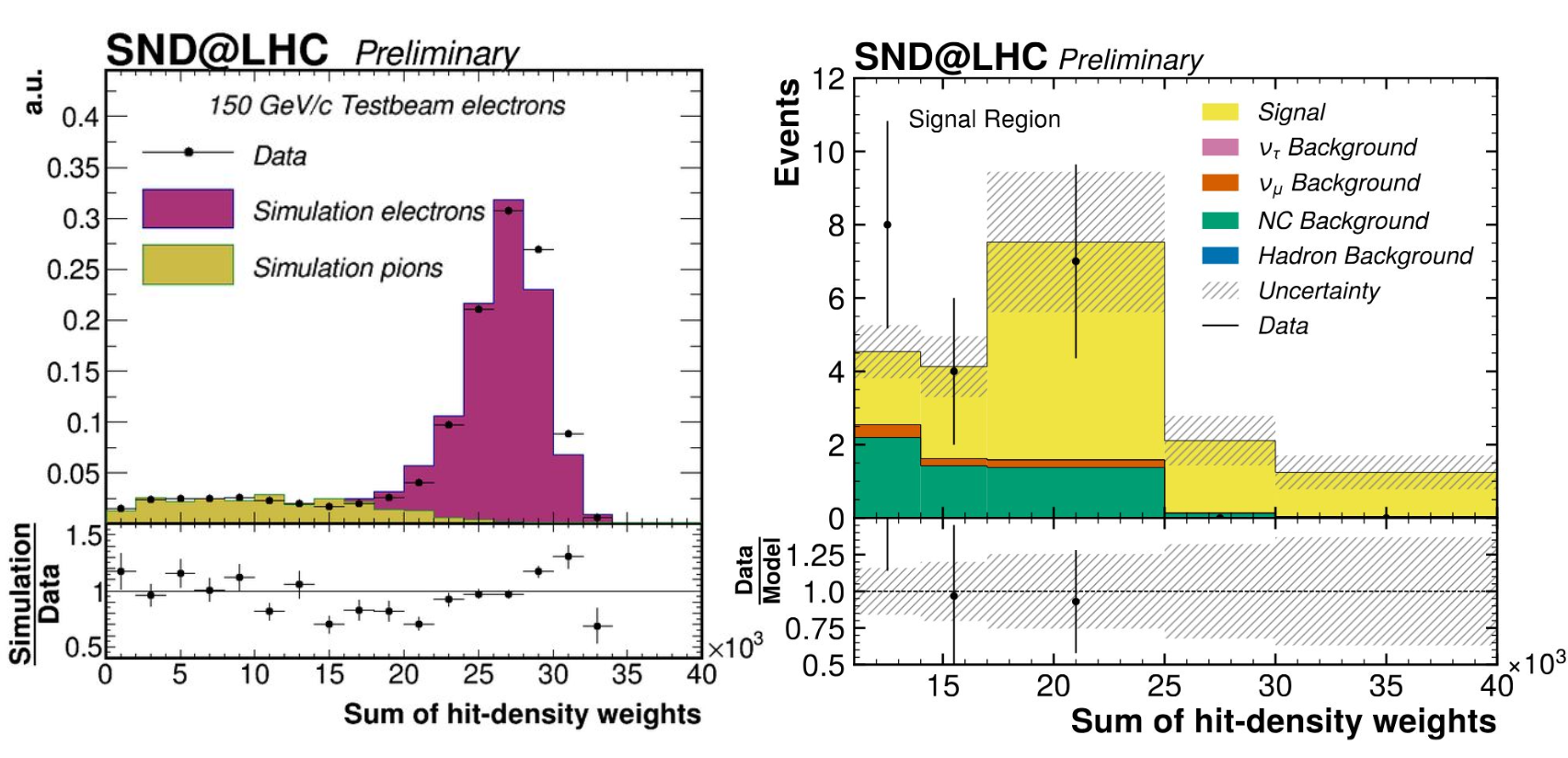}
\caption{Distributions of the summed hit-density weights in the target used in the zero muon analysis for (left) test-beam data samples and (right) the 2022--2024 data sample corresponding to an integrated luminosity of 170.6~fb$^{-1}$~\cite{SNDICHEP26}.}
\label{fig:SND@LHC-new}
\end{figure}

SND@LHC has also presented an updated analysis of the zero muon event sample, including the 2024 dataset and following the strategy of the published analysis~\cite{SNDLHC:2024qqb}. Using the 2022--2024 data sample, 19 events are observed, compared with an expectation of $19\pm5$ events, consisting of $13\pm5$ signal and $6\pm1$ background events. The observed significance of the $\nu_e$ CC contribution is 2.7 standard deviations in the binned likelihood analysis and 3.6 standard deviations in the corresponding unbinned analysis~\cite{SNDICHEP26}. The preliminary result is shown in Fig.~\ref{fig:SND@LHC-new}, using the same search variable as in the previous analysis, namely a hit-density weight variable built from the local density of SciFi hits within 1~cm in a given station. Further ongoing activities include a study of muon trident interactions in the upstream rock in front of the experiment, using the channel $\mu+N \rightarrow \mu\mu\mu + N$, as well as the search for an anomalous $\nu_{\tau}$ magnetic moment, based on events with a single electromagnetic shower and no hadronic activity. For this latter study, a preliminary sensitivity of a few $\times 10^{-6}\,\mu_B$ is expected with the Run~3 data, improving to $10^{-7}\,\mu_B$ at the HL-LHC~\cite{SNDICHEP26}.

In addition to the neutrino interaction analyses, SND@LHC has performed dedicated studies of the muon background in TI18. These measurements are important because long range muons produced at the ATLAS interaction point constitute the dominant background for neutrino searches and also determine the exposure rate of the emulsion target. One study measured the muon flux in the central $31\times31$~cm$^2$ fiducial region using 2023--2025 LHC collision data, combining independent measurements from the SciFi tracker and the downstream muon system. It reported the flux evolution for the different data taking periods and found agreement with Monte Carlo predictions within the quoted uncertainties~\cite{SNDLHC:2026muf}. A complementary study provided a broader characterisation of the Run~3 muon background over the full ECC acceptance, comparing the measured rates and angular distributions with the FLUKA-based simulation chain for the different LHC optics and crossing configurations~\cite{SNDLHC:2026yqa}. At a reference instantaneous luminosity of $2\times10^{34}$~cm$^{-2}$s$^{-1}$, the measured rates were 557, 1154 and 799~Hz for the 2022--2023, 2024 and 2025 configurations, respectively, with simulation and data agreeing at the level of about 10--15\%. The increase observed in 2024 was associated with the reverse-polarity LHC optics, while the remaining enhancement in 2025 was linked to the horizontal crossing scheme and diffractive proton losses in the dispersion suppressor region. These studies validate the background model used in Run~3 analyses and provide important input for the SND@HL-LHC upgrade.

The LHC Run~3 concluded in June 2026. By the end of Run~3, both the FASER and SND@LHC experiments had accumulated substantially larger neutrino datasets than those used in their first published measurements, enabling more differential studies and improved constraints on the forward neutrino flux. SND@LHC, which is located off-axis, provides complementary coverage in pseudorapidity. The installation of a third veto layer at the end of 2023 improved the background veto coverage and is expected to increase the neutrino acceptance by about a factor of 1.5 for the 2024--2026 data analysis. The predicted numbers of neutrino DIS interactions in SND@LHC for Run~3 are summarised in Table~\ref{tab:neutrino_yields1}. Since FASER is located on the collision axis, where the far-forward neutrino flux is largest, the Run~3 predictions indicated a larger neutrino sample for FASER than for SND@LHC, typically by roughly a factor of 3--6 depending on the neutrino flavour, detector acceptance, luminosity normalisation, and flux model.

\begin{table}[b]
\centering
\begin{tabular}{c  c  c}
\toprule
Flavour & DIS-CC  & DIS-NC  \\
\midrule
$\nu_{\mu} + \bar{\nu}_{\mu} $ &
1575  & 508 \\
$\nu_e + \bar{\nu}_e $ & 484 &  161 \\
$\nu_{\tau} + \bar{\nu}_{\tau}$ & 37 & 19 \\
\midrule
Total  &   2096   & 688 \\
\bottomrule
\end{tabular}
\caption{Predicted number of neutrino DIS interactions in SND@LHC for Run~3 with an anticipated luminosity of 310~fb$^{-1}$. The values are based on the flux and generator assumptions used in the SND@LHC study~\cite{SNDMoriond26}.} 
\label{tab:neutrino_yields1}
\end{table}

\section{Prospects for the HL-LHC}
\label{sec:prospects}
The prospect of turning the LHC into a neutrino source has now moved from early concepts to a more concrete programme for the HL-LHC era. This development has been outlined in a recent road map~\cite{Graverini:2024ynx} and in studies for a Forward Physics Facility~\cite{Anchordoqui:2021ghd,Feng:2022inv}. One approach is to extend the current experiments, located in the tunnels about 480~m from the ATLAS interaction point, for at least Run~4, the first run of HL-LHC operation starting around 2030, and possibly over the full HL-LHC running period. This would allow larger samples to be collected and would open more analysis channels, in particular those related to $\nu_{\tau}$ studies. The other approach is to build a new underground facility in the line of sight of the proton beam, able to host larger detectors and provide additional shielding and services. Such a facility would also make it easier to operate several detectors using complementary technologies in one place. This can be instrumental in understanding common backgrounds using different techniques, allowing systematic cross checks, and ultimately combining information from these possible experiments to extend the physics reach. In this section, the upgrade proposals based on the present neutrino experiment locations planned for Run~4 and beyond are summarised first, followed by the Forward Physics Facility concept and other alternative experimental ideas.

\subsection{Upgrade proposals for LHC Run~4 and beyond}
Two upgrade proposals have been discussed for the neutrino programme at the ATLAS interaction point for Run~4 and the later HL-LHC running, building on the Run~3 tunnel experiments. In both cases the main goals are similar. The event yield is driven by the target mass and geometric acceptance, while the dominant backgrounds are driven by through going muons and muon-induced secondaries in the far-forward region. The proposed upgrades therefore focus on increasing acceptance and target mass as much as the experimental site allows, and on improving the veto system and reconstruction to keep muon-induced backgrounds and detector systematic uncertainties under control. 

Notably, the use of the standard emulsion detector technique in these upgrades is expected to be challenging. Due to the expected large increase in LHC luminosity, as shown in Fig.~\ref{fig:LHC}, the emulsions would need to be exchanged every few weeks, which could become impractical and strongly affect LHC running. FASER has an ongoing R$\&$D programme to assess what luminosity could be collected with a single emulsion box under HL-LHC conditions. Possible mitigation strategies are being examined, such as remotely tilting the emulsion detector to spread the muon background over angular space, as well as improvements in emulsion chemical composition, scanning, and reconstruction. At the same time, several active target concepts are also being studied as alternatives to the current emulsion-based approach, with some expected loss of vertex resolution power.

\begin{figure}
\centering
\includegraphics[width=0.49\textwidth]{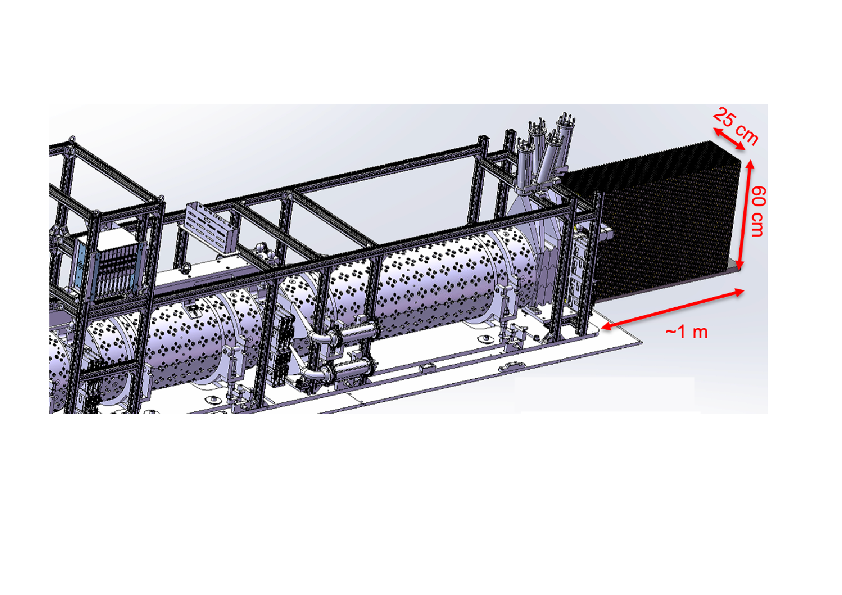}
\includegraphics[width=0.44\textwidth]{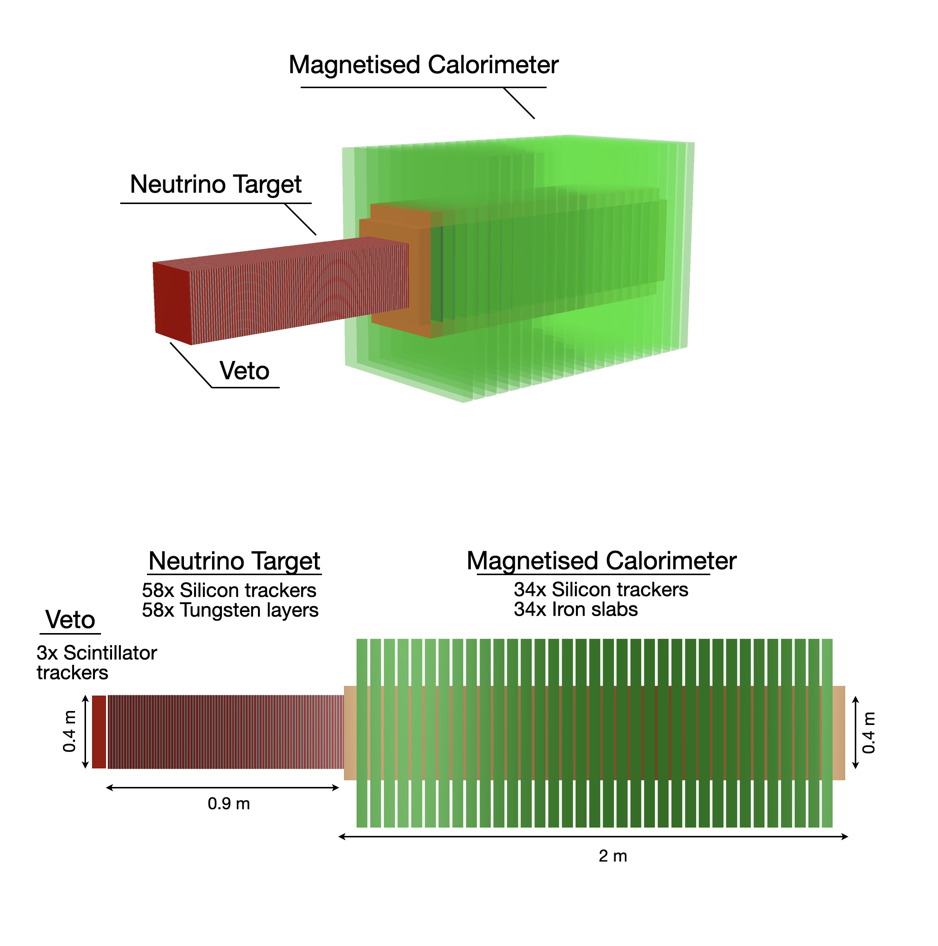}
\caption{Schematic views of two upgrade concepts discussed for the ATLAS interaction point collider neutrino programme. Left: reference on-axis upgraded FASER neutrino detector concept with an active tungsten target and readout, designed to fit within the TI12/UJ12 tunnel constraints. Right: upgraded SND@HL-LHC concept showing the neutrino target, veto system, and magnetised calorimeter for charge and momentum measurement. Adapted from Refs.~\cite{FASER:2025myb,SNDLHC:2026why} (CC BY 4.0).}
\label{fig:run4_upgrades}
\end{figure}

\begin{itemize}  
  \item \textbf{FASER and FASER$\nu$:} The upgrade concept for FASER is illustrated in Fig.~\ref{fig:run4_upgrades} (left). An upgraded FASER neutrino detector has been proposed as input to the European Particle Physics Strategy (ESPP) Update~\cite{ESPP:2020,ESPP:2026Recommendations}, with the aim of operating in Run~4 and possibly beyond~\cite{FASER:2025myb}. The reference on-axis design includes a tungsten target and active readout, within the constraints of the current tunnel site dimensions, and is used to estimate event yields and expected precision in Run~4 and the later HL-LHC runs. The main idea is to gain more statistics and improve event reconstruction, so that the programme can go beyond the inclusive $\nu_\mu$ CC channel and access more flavours and interaction topologies. The physics programme discussed includes more precise TeV energy CC cross section measurements, first precise measurements of additional flavours and antineutrino components, and extended samples that can separate CC and NC contributions and constrain the forward neutrino flux. Several detector options are presently discussed, including high-granularity scintillator-based tracking calorimeters, additional silicon tracking layers, and advanced emulsion-based configurations for more exclusive reconstruction. FASER has already been approved to extend operation into Run~4 with a purely passive tungsten neutrino target.
  
  In addition to the on-axis neutrino detector, a scintillator-based off-axis detector located downstream of FASER is also being considered. Two prototype detectors for this concept were installed for the 2026 data taking period.
  
  \item \textbf{SND@HL-LHC:} The upgrade concept for SND@HL-LHC is illustrated in Fig.~\ref{fig:run4_upgrades} (right). An upgraded SND@LHC programme at the ATLAS interaction point has been studied and has evolved into an SND@HL-LHC Technical Proposal for Run~4~\cite{Abbaneo:SNDHLHC:LHCC-P-026}, with further physics reach studies and installation scenarios discussed in~\cite{SNDLHC:2026why}. The upgrade concept keeps an off-axis configuration and studies two installation scenarios, Baseline and Extended, in TI18. The ``Extended'' option proposes a trench to allow the detector to be lowered and shifted transversely, improving the overlap with the forward neutrino flux and increasing the total neutrino interaction rate by about a factor of five. Expected event rates for both configurations are presented in Table~\ref{tab:neutrino_yields2}. The upgraded detector replaces the emulsion target with a fully electronic design based on silicon tracking detectors to cope with the higher muon rates expected at the HL-LHC, and introduces a magnetised tracking calorimeter to measure muon charge and momentum, allowing neutrino and antineutrino interactions to be separated. The physics reach discussed in this study includes separate $\nu_\mu$ and $\bar{\nu}_\mu$ cross section measurements, constraints on PDFs at very small $x$, tests of lepton flavour universality, additional opportunities such as the potential combination of SND@LHC data with information from ATLAS activity for certain selected events, and sensitivity to tau antineutrino interactions. The technical proposal has been approved and preparation for the Run~4 detector is expected to start at the end of 2026.
\end{itemize}

\begin{table}[b]
\centering
\begin{tabular}{c | c c | c c}
\toprule
& \multicolumn{2}{c|}{\textbf{Baseline}} & \multicolumn{2}{c}{\textbf{Extended}} \\
\midrule
Flavour & Target & Target+HCAL & Target & Target+HCAL \\
\midrule
$\nu_\mu$         & 1.1$\times 10^4$ & 1.7$\times 10^4$ & 5.4$\times 10^4$  & 7.5$\times 10^4$  \\
$\bar{\nu}_\mu$   & 3.6$\times 10^3$ & 6.1$\times 10^3$ & 1.7$\times 10^4$  & 2.3$\times 10^4$  \\
$\nu_e$           & 1.7$\times 10^3$ & 2.8$\times 10^3$ & 7.3$\times 10^3$ & 9.7$\times 10^3$ \\
$\bar{\nu}_e$     & 7.6$\times 10^2$ & 1.2$\times 10^3$ & 2.7$\times 10^3$ & 4.4$\times 10^3$ \\
$\nu_\tau$        & 1.0$\times 10^2$ & 1.7$\times 10^2$ & 2.4$\times 10^2$ & 3.4$\times 10^2$ \\
$\bar{\nu}_\tau$  & 5.0$\times 10^1$ & 8.9$\times 10^1$ & 1.2$\times 10^2$ & 1.7$\times 10^2$ \\
\midrule
Total             & 1.7$\times 10^4$ & 2.7$\times 10^4$ & 8.1$\times 10^4$ & 1.1$\times 10^5$ \\
\bottomrule
\end{tabular}
\caption{
Predicted number of neutrino CC DIS interactions for SND@LHC, as obtained from the \textsc{DPMJET}~\cite{Roesler:2000he} (light meson decays) and \textsc{POWHEG}~\cite{Alioli:2010xd} (charm hadron decays) simulations assuming 3000~fb$^{-1}$ of integrated luminosity for the HL-LHC. Both the Baseline and Extended configurations are shown, distinguishing interactions in the target and in the combined target+HCAL volume~\cite{SNDLHC:2026why}.}
\label{tab:neutrino_yields2}
\end{table}

\subsection{Forward Physics Facility concepts}
\begin{figure}
\centering
\includegraphics[width=0.60\textwidth]{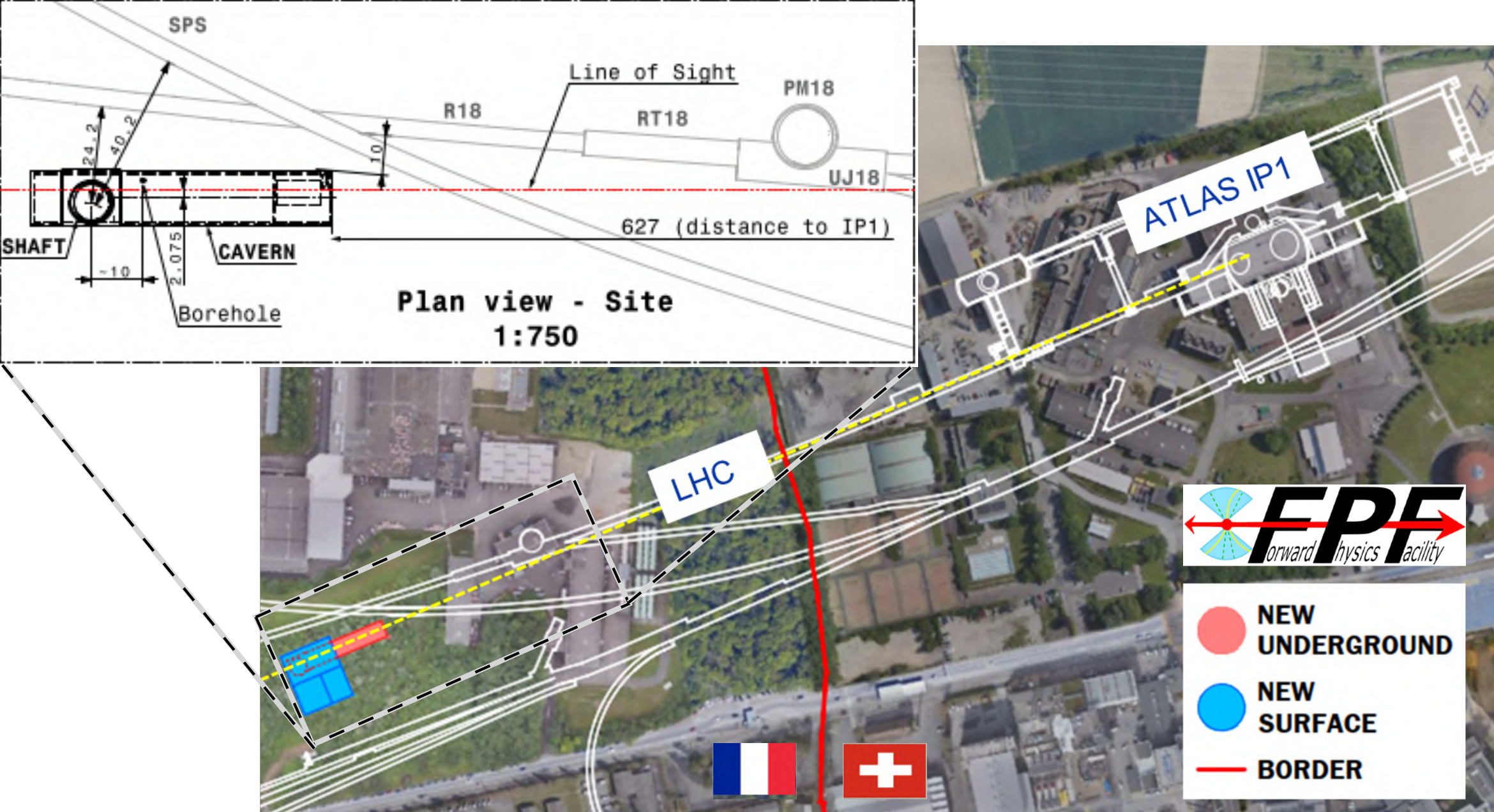}
\caption{Site and line of sight sketch for the proposed Forward Physics Facility (FPF) cavern location. Adapted from Ref.~\cite{Anchordoqui:2025fpf} (CC BY 4.0).}
\label{fig:fpf_location}
\end{figure}

\begin{figure}
\centering
\includegraphics[width=0.50\textwidth]{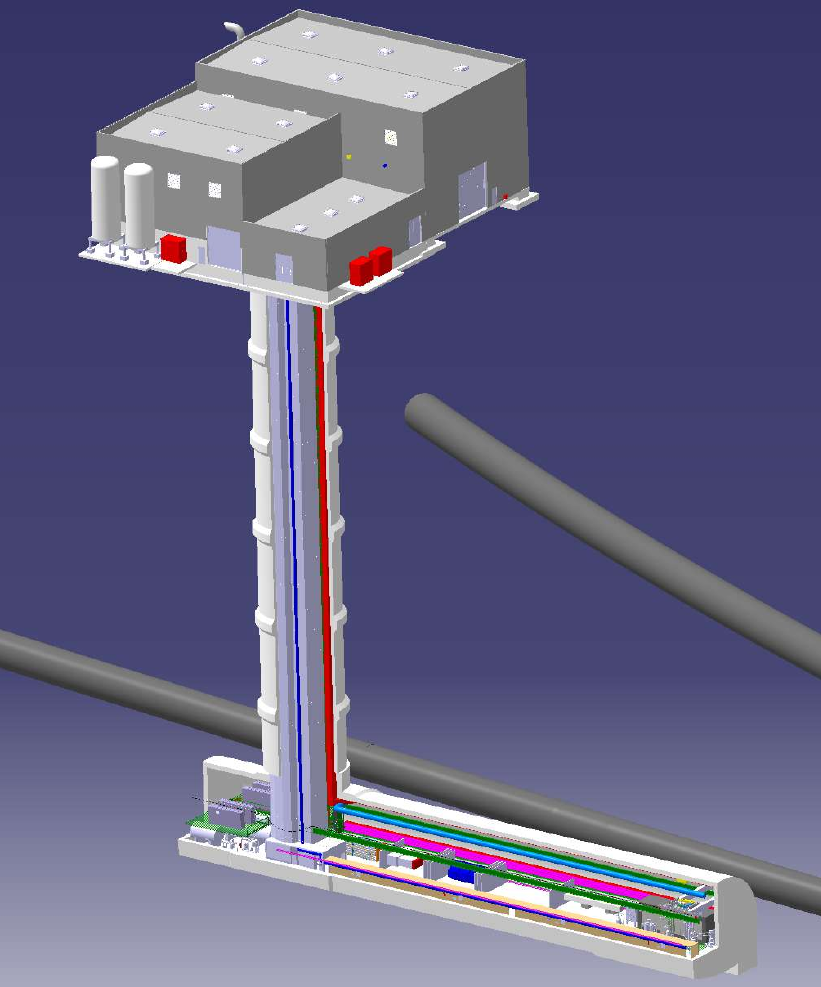}
\caption{3D view of the surface building, shaft, and underground cavern for the proposed FPF. Adapted from Ref.~\cite{Anchordoqui:2025fpf} (CC BY 4.0).}
\label{fig:fpf_3d}
\end{figure}

\begin{figure}
\centering
\includegraphics[width=0.80\textwidth]{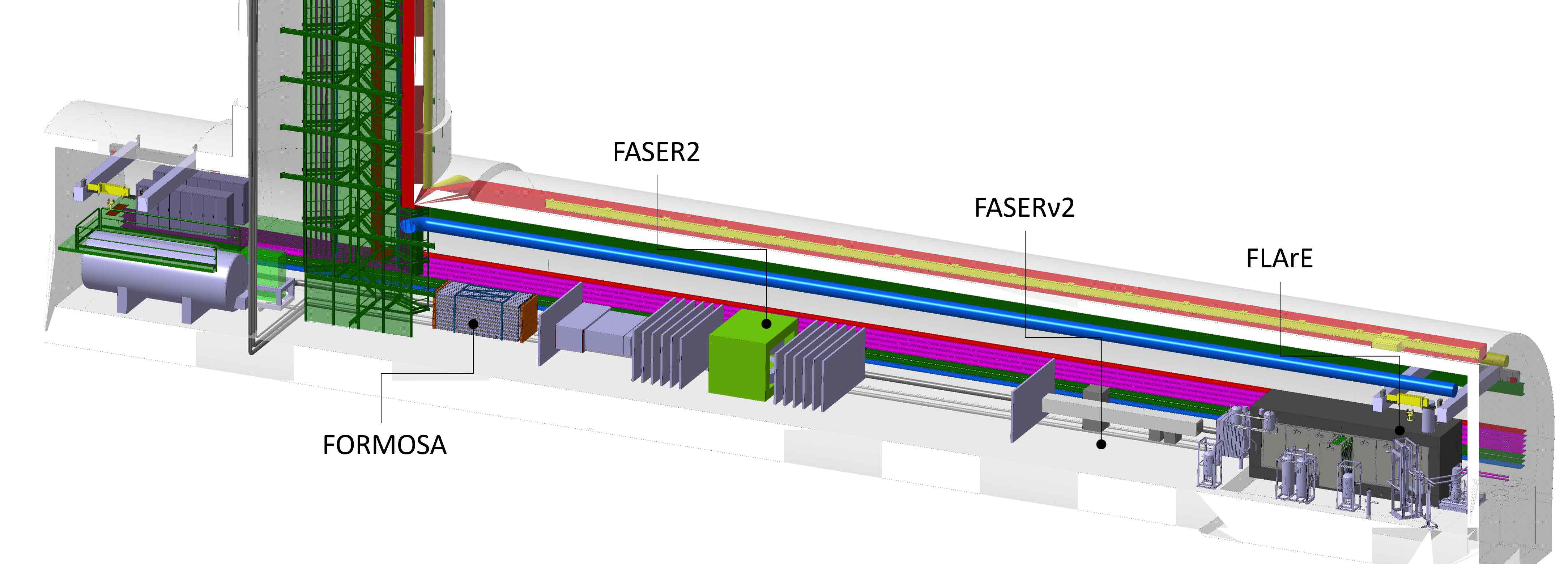}
\caption{Layout of the proposed FPF cavern with the placement of the four detector concepts (FASER2, FASER$\nu$2, FLArE, and FORMOSA). Adapted from Ref.~\cite{Anchordoqui:2025fpf} (CC BY 4.0).}
\label{fig:fpf_baseline}
\end{figure}

The Forward Physics Facility (FPF) is a proposal for a dedicated large far-forward underground cavern for the HL-LHC era, with large space, proper shielding, and services to host several experiments in one location. The main motivation is simple. The presently used tunnel sites allow the installation of compact detectors, but they are strongly limited in detector size, access, and shielding options. A facility-style cavern is meant to remove these constraints and support larger detector configurations with more flexible layouts and stronger background mitigation possibilities~\cite{Anchordoqui:2021ghd,Feng:2022inv}. The proposed site and facility layout are illustrated in Figs.~\ref{fig:fpf_location} and \ref{fig:fpf_3d}. The proposed cavern has a length of 75~m, with 65~m available for experiments, and a width of 12~m, at a depth of about 100~m underground. The distance from the cavern to the ATLAS interaction point is 627~m. The current FPF experimental concept consists of four complementary detector projects, as shown in Fig.~\ref{fig:fpf_baseline}, each targeting a complementary part of the forward physics programme.

\begin{itemize}
  \item \textbf{FASER2:} a large on-axis magnetic tracking spectrometer with electromagnetic and hadronic calorimetry and muon detectors, designed to reconstruct forward hadrons, electrons, muons and photons, and including a decay volume for dark sector particle searches of a few orders of magnitude larger than that of FASER. It provides the charged particle, lepton and photon measurements needed for long-lived particle (LLP) searches and also supports neutrino studies through muon tagging and kinematic reconstruction in the forward direction. The detector layout and expected performance of FASER2 have been studied in detail in a dedicated design note~\cite{Salin:FASER2:PBCNOTE2025-005}.

  \item \textbf{FASER$\nu$2:} a large neutrino emulsion--tungsten target upstream of the FASER spectrometer to collect high statistics interaction samples. It targets precision measurements of TeV scale neutrino cross sections and flavour samples, with sensitivity to $\nu_e$ and $\nu_\tau$ through high resolution vertexing and lepton identification.
  
  \item \textbf{FLArE:} a liquid-argon time projection chamber (LArTPC) detector concept that provides a massive target and rich event information for neutrino interactions. LArTPCs are like electronic ``bubble chambers'', giving high quality information on the full interaction, and have been deployed successfully in recent years in several accelerator-based lower energy neutrino experiments. FLArE is aimed at high statistics neutrino samples with improved final state reconstruction, including hadronic activity, which is important for differential measurements, NC studies, and nuclear effects in the TeV energy range.

  \item \textbf{FORMOSA:} a dedicated detector concept for detecting millicharged particles in the far-forward region. It is optimised for sensitivity to tiny ionisation signals from feebly interacting charged particles that can be produced in meson decays or Drell--Yan processes at the LHC and travel in the forward direction. This detector consists primarily of scintillator bars and is based on the milliQan detector concept~\cite{milliQan:2021lne}. A small FORMOSA demonstrator was installed downstream of FASER during the 2024--2025 running period to validate the detector concept. A sensitivity for 
  millicharges to below $q=10^{-3}$ should be reachable.
\end{itemize}

Overall, the FPF is proposed as a way to push the forward physics programme beyond what is possible in the current LHC tunnels. For neutrino physics, the focus is on large samples at TeV energies, with event yields about a factor of 5--20 larger than what could be expected from upgrades of the experiments in the tunnel. This obviously comes at a price, namely a new cavern, new large experiments, and more complexity in organisation. But the physics programme is very broad and unprecedented. It covers a wide range of neutrino and electroweak measurements, opens a new window into QCD with the FPF effectively acting as a neutrino--ion collider, analogous and complementary to the electron--ion collider~\cite{Deshpande:2002em,Blednykh:2026gii} expected to start operation at BNL around 2032, and provides new precision insight into the parton structure of the proton. The data would also have strong connections to astroparticle physics, potentially giving new insight into, for example, the muon puzzle, strangeness enhancement, and the charm backgrounds from atmospheric showers. The facility also supports a broad BSM programme, including searches for feebly interacting particles, dark matter, and sterile neutrinos.

Recent documents have summarised the physics opportunities and the conceptual design, including tunnel site and layout options, shielding and services, and the experimental constraints from muon-induced backgrounds~\cite{Adhikary:2024nlv,Anchordoqui:2025fpf}. To date, geological and expected background studies have been carried out, as well as studies of possible interference with the LHC during cavern preparation. These studies have not shown any showstoppers, but a decision to proceed would need to be made in a timely way for the FPF to benefit from the full HL-LHC running period.

\subsection{Alternative experimental ideas}
Besides the upgrade projects for the current tunnel experiments and the FPF proposal, which is still to be decided, several other detector concepts have been discussed. These target not only collider neutrinos from $pp$ collisions, but also neutrinos from atmospheric and astrophysical sources. The main idea is to choose different geometries and locations to exploit specific advantages. Some concepts stay close to the interaction point and use existing LHC infrastructure or an existing detector. Others move to longer baselines and compensate the reduced flux with a much larger detector volume. Representative ideas and proposals are summarised below.

\begin{figure}
\centering
\includegraphics[width=0.60\textwidth]{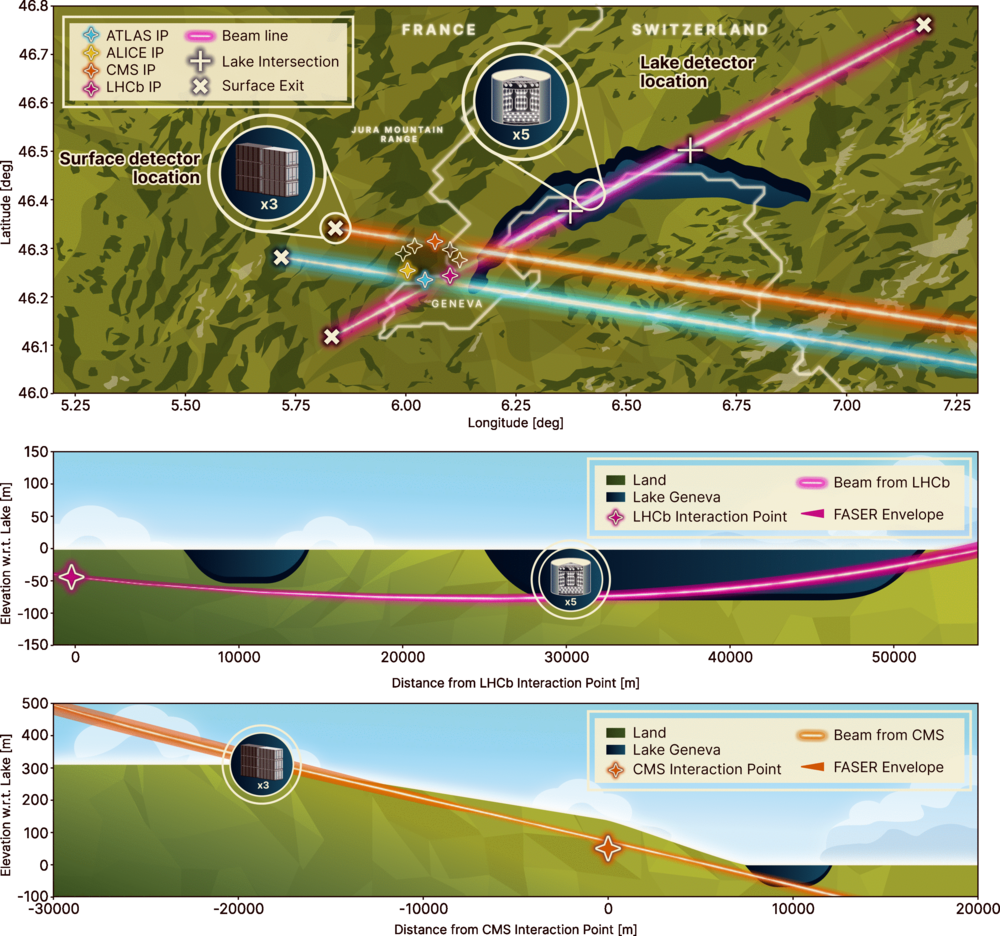}
\caption{Schematic overview of the SINE and UNDINE concepts, showing detector locations on the surface and under the lake, and the corresponding beam geometries. The shaded envelope shown around each beamline corresponds to the profile of the neutrino beam passing through the FASER detector. Adapted from Ref.~\cite{Kamp:2025phs} (CC BY 4.0).}
\label{fig:sine_undine}
\end{figure}

\begin{itemize}
  \item \textbf{Medium baseline detectors near the surface:}
  There is an alternative way to hunt for forward-produced neutrinos: simply let them come to the surface. Given the inclination angle of the LHC machine with respect to the surface in the Geneva area and the fact that the earth is a sphere, the forward neutrinos produced at the LHC will eventually reach the surface somewhere. This has been studied by several groups. Studies of forward neutrino propagation through the surrounding rock and molasse in the Geneva basin locate where the LHC neutrino flux can exit to the surface, giving concrete candidate sites for possible surface level detectors~\cite{Ariga:2025gtj}. The key point is geometry. The far-forward neutrino flux is strongly collimated, so only specific directions and baselines give a significant flux at the surface, and the detector location becomes part of the concept. Building on this, the SINE and UNDINE proposals consider medium baseline detectors with much larger target masses, including a surface scintillator setup and a Geneva lake-based water volume~\cite{Kamp:2025phs}. In these setups the neutrino flux is smaller than for near interaction point options, but the loss is compensated by the potentially large detector mass, so large samples can still be collected. The experimental challenges are also different. Cosmic ray backgrounds and event timing become more important at the surface, while accelerator backgrounds are negligible and detector access and operation may be generally easier than in the tunnel or cavern. The basic geometry and some example detector locations are illustrated in Fig.~\ref{fig:sine_undine}. Note that these detector locations are not presently on CERN owned land, which may complicate the matter. This scenario was also studied in Ref.~\cite{Ariga:2025gtj}, where experimental challenges were identified for such a setup. It was concluded there that, at present, the physics potential of surface level detectors is limited compared with detectors closer to the interaction point, including the proposed Forward Physics Facility. The jury on what is optimal/feasible to realise is therefore still out.
  
  \item \textbf{Non-forward neutrinos in the LHC central detectors:} 
  A different approach is to use one of the LHC general-purpose detectors as an active neutrino target. This idea has been discussed for CMS, in particular using the upgraded high-granularity calorimeter (HGCAL) at the HL-LHC~\cite{Foldenauer:2021gkm}. The dense detector material provides a large effective target mass, while the fine spatial and timing resolution can be used to identify displaced signatures associated with neutrino interactions. The main challenge is separating rare neutrino-induced events from the large background of $pp$ collision products. A first study demonstrates the potential, based on simple cuts, but the expected statistics are low, with about four events and a potential signal-to-background sensitivity $S/\sqrt{B}$ of about 2, using the full HL-LHC luminosity. A detailed prospective reach, for example using more sophisticated analysis techniques such as machine learning, has not been established yet.
  
  Another proposal using the existing central LHC detectors has been studied in Ref.~\cite{GarciaSoto:2025shift}. In this case, SHIFT@LHC, a proposed gaseous fixed target installed in the LHC tunnel, would provide a novel source of detectable neutrinos. Based on simulations of proton--gas collisions, hadron propagation, and neutrino interactions, about $\mathcal{O}(10^4)$ muon neutrino and $\mathcal{O}(10^3)$ electron neutrino interactions, with energies between 20~GeV and 1~TeV, are expected in the CMS and ATLAS detectors with about 1\% of the LHC Run~4 integrated luminosity. This configuration would provide access to hadron production in the very forward region, which is directly relevant to atmospheric neutrino experiments.
  
  \item \textbf{Atmospheric and astrophysical neutrinos in ATLAS:}
  Several studies have discussed using the ATLAS detector itself for neutrino measurements, mainly because of its large calorimeter mass and muon system. An early idea focused on atmospheric neutrino oscillations, especially $\nu_\mu$ events~\cite{Kopp:2007ne}. Two event classes were considered, including upward going muons from $\nu_\mu$ interactions in the rock, and contained $\nu_\mu$ CC events with a visible hadronic shower in the calorimeter. Such searches are most sensitive during periods without $pp$ collisions, when the detector environment is quieter and the trigger can focus on non-beam activity. Of course, one would have to keep the detector switched on during these non-beam periods.
    
  More recent papers revisit ATLAS in a different context. One example concerns readiness for detecting high energy neutrinos from a Galactic supernova~\cite{Wen:2023ijf}. The focus is on a possible high energy neutrino component that could produce muons or showers in ATLAS, rather than on the more generally expected MeV energy neutrino burst. Another example is to use the ATLAS magnetic field and muon charge measurement to access the atmospheric $\nu_\mu/\bar{\nu}_\mu$ flux ratio~\cite{Ghosh:2024ryg}. In this case, muons with measured charge provide a handle to separate neutrino and antineutrino contributions.
\end{itemize}
 
\section{Physics opportunities and outlook}
\label{sec:physics}
The first observation of neutrino interactions at the LHC in Run~3 has opened a new direction for neutrino physics. These measurements show that collider neutrinos can be studied experimentally, even in the difficult environment of a hadron collider. Although the present data sample is still limited, the first results already demonstrate the physics potential of this programme. In the HL-LHC era, with much larger integrated luminosity and possible new detector configurations, collider neutrino studies can move from first measurements to a broader precision measurement programme. Below we mention the main physics topics that can be addressed with these collected datasets, in particular with the already approved experiments for HL-LHC running starting around 2030.

\paragraph{Neutrino cross sections at TeV energies.}
One of the main physics goals is the measurement of neutrino interaction cross sections at TeV energies. This energy range is beyond that of traditional accelerator-based neutrino experiments. With larger data samples at the HL-LHC, it will be possible to measure precise CC and NC processes and cross sections, and to study their dependence on energy, rapidity and event topology. Such measurements would extend our knowledge of neutrino--nucleon scattering to a new kinematic region and provide useful input for high energy neutrino physics. These extended cross section measurements as a function of neutrino energy will also be very important for present and future large neutrino experiments looking at the sky for high energy atmospheric and astrophysical neutrinos, such as the IceCube Neutrino Observatory~\cite{IceCube:2021rpz} at the South Pole and KM3NeT~\cite{KM3NeT:2025npi} in the Mediterranean Sea. In addition, in this new high energy regime of accelerator controlled neutrino interactions, precision cross section measurements will also allow the data to be tested for the presence of non-standard neutrino interactions (NSI).

\paragraph{Neutrino flavours and event topologies.}
The forward neutrino flux at the LHC contains all neutrino flavours. In particular, it includes a significant contribution of tau neutrinos from heavy flavour meson decays. This makes the LHC an interesting place to study $\nu_\tau$ interactions, which are still much less explored than $\nu_e$ and $\nu_\mu$ interactions. With higher statistics and improved detector performance, future experiments may also study different classes of neutrino events in more detail. Sensitivity to antineutrino interactions will also be possible. Efficient flavour identification will allow comparisons between different neutrino flavour rates, giving sensitivity to tests of lepton flavour universality in a new energy range, in particular between electron and tau flavours, which are least contaminated by non-heavy flavour production processes.

\paragraph{Neutrinos from charm hadrons.}
As mentioned before, at high energies $\nu_e$ and $\nu_\tau$ neutrinos mainly stem from decays of charm hadrons. Hence, from these neutrino rates one can extract information on charm quark production in $pp$ collisions and, through pQCD, obtain information on the gluon distribution in the proton at small Bjorken-$x$ values down to $10^{-4}$ and relevant momentum transfer scales. When included in global fits, this can considerably reduce the uncertainty on the gluon PDFs in the phase space region relevant for LHC energies and possible future higher energy colliders. At the same time, this measurement will also be of strong interest for experiments such as IceCube in estimating the atmospheric neutrino background for measurements of astrophysical neutrino energy spectra, in the region of $10^4-10^6$ GeV, where the poorly known neutrino contribution from charm mesons produced in atmospheric showers presently leads to a large uncertainty~\cite{Bustamante:2017xuy}.

\paragraph{Forward hadron production and neutrino flux.}
The neutrino flux at the LHC depends directly on the production of hadrons in the forward region. For this reason, collider neutrino measurements are closely connected to QCD particle production in the forward region, which for light particles is described by phenomenological models that need to be tuned to experimental data. The production of hadrons containing charm quarks is important for the $\nu_\tau$ component of the flux. Measurements of neutrino rates and spectra can therefore help in testing and constraining models of forward particle production. In this way, collider neutrino experiments are sensitive not only to neutrino interactions, but also to hadron production in a phase space region that is still not well constrained. First results have been released by FASER on implications of their neutrino measurements for forward hadron production in $pp$ collisions at the LHC~\cite{FASER:2025sge}, based on the first two years of data taking. The measured neutrino fluxes are compared with predictions from recent hadronic interaction models, which are found to be generally consistent with the FASER measurements, although some notable discrepancies are observed and will be followed up with more data.

\paragraph{Neutrino--nucleus interactions.}
Collider neutrinos also provide access to neutrino--nucleus scattering at high energies. This is interesting because, for example, nuclear effects can be studied in a kinematic region different from that covered by current and past lower energy fixed target experiments. In particular, detailed neutrino--nucleus interaction measurements with neutrinos at LHC energies cover a kinematic region similar to that expected from electron--nucleus collisions at the Electron-Ion Collider (EIC), which is expected to start collecting data in the 2030s. Hence, the availability of far-forward neutrino detectors at the LHC effectively provides CERN with a charged-current counterpart to the EIC~\cite{AbdulKhalek:2021gbh}. Measurements in this regime will improve our understanding of the role of the nucleus in neutrino interactions~\cite{CruzMartinez:2023sdv}. Such information is also relevant for the modelling of high energy neutrino interactions in astroparticle physics.

\paragraph{Searches for physics beyond the Standard Model.}
Several detector concepts built or proposed for collider neutrino studies can also be used to search for new physics signatures, such as new weakly coupled particles. Examples include heavy neutral leptons 
(HNLs), dark-sector particles, axion-like particles, and other long-lived states. The forward direction is especially interesting because many such particles can be produced at small angles with respect to the beam axis, either in hadron decays or via Drell--Yan processes. As a result, experiments designed for neutrino physics may also contribute to a broader programme of searches for BSM physics. The FASER detector was in fact designed from the start to search for new light, weakly interacting particles, with its broader BSM physics prospects discussed in Ref.~\cite{FASER:2018eoc}. It has already produced world-best limits with early Run~3 data in searches for dark photons~\cite{FASER:2023tle}, with updated results now available~\cite{Conf-FASER-CONF-2026-001}, as well as searches for axion-like particles~\cite{FASER:2024bbl}. Prospects for SND@LHC are given in Refs.~\cite{SNDLHC:2026why,Boyarsky:2021moj}, in particular for scattering processes induced by light dark sector particles produced in the $pp$ interactions. The FPF has the widest reach for different species of new-particle searches, as detailed in Ref.~\cite{Anchordoqui:2025fpf} and discussed in the previous section.

\paragraph{Complementarity with other neutrino experiments.}
Collider neutrino measurements are complementary to those from other neutrino experiments. Traditional accelerator experiments usually work at lower energies and with well-controlled beams, while atmospheric and astrophysical neutrino experiments probe much higher energies but with less direct control of the source. The LHC provides a different regime, including neutrinos at high energies, produced in $pp$ collisions, and concentrated in the forward direction. For this reason, collider neutrino measurements will connect different parts of neutrino physics and provide useful information for the interpretation of high energy neutrino data.

In summary, the HL-LHC offers the possibility to develop collider neutrino physics into a broader experimental programme. With larger data samples, improved detectors, and a better understanding of the forward flux, future measurements can address not only neutrino cross sections, but also flavour physics, neutrino--nucleus interactions, forward hadron production, and searches for new physics. Collider neutrinos therefore provide a new opportunity to study both the Standard Model and possible phenomena beyond it.

\section{Conclusion}
\label{sec:conclude}
The observation of neutrino interactions at the LHC has turned a long-standing idea into an experimental reality. During the LHC Run~3, the FASER and SND@LHC experiments showed that collider neutrinos can be detected and studied in the difficult environment of a hadron collider. They have reported the first observations and first measurements in this new regime. These results establish the LHC as a source of high energy neutrinos, open a new direction for neutrino physics, and connect the field with wider topics such as astroparticle physics, precision QCD and BSM searches.

In the HL-LHC era, this programme is expected to grow further. Larger data samples and improved detector concepts will allow more detailed studies of neutrino interactions at TeV energies and in the forward region. Collider neutrino experiments can therefore develop into a broader physics programme connecting neutrino interactions, forward hadron production, and searches for physics beyond the Standard Model. The LHC is a unique machine, and it will likely be a long time before another facility of this kind becomes available. Hence we should make sure that we do not miss the opportunity to explore this exciting neutrino and beyond programme during its lifetime.

The collider neutrino programme has also triggered studies for possible future colliders, where neutrino detectors could be included in the experimental design from the beginning. Examples are studies for the FCC$_{hh}$~\cite{Abraham:2024fcc} and for a future muon collider~\cite{Kling:2026NSI,Adhikary:2025npd,Marzocca:2025qmk}. In particular, a muon collider would produce a very large neutrino flux from beam muon decays, potentially reaching interaction rates of order $10^5$ interactions per kg per year. These studies show that collider neutrino physics could become an important part of the physics programme of future high energy facilities.

\ack{We would like to thank Giovanni De Lellis, Jamie Boyd, and Felix Kling for very useful discussions during the preparation of this manuscript. This work was supported by the Reinventing University Program of the Office of the Permanent Secretary, Ministry of Higher Education, Science, Research and Innovation (MHESI), Thailand. This work also received support from the CU Power Grant, The Second Century Fund (C2F), Chulalongkorn University.}

\bibliography{reference}

@article{FASER:2024ykc,
    author = "Mammen Abraham, Roshan and others",
    collaboration = "FASER",
    title = "{Neutrino rate predictions for FASER}",
    eprint = "2402.13318",
    archivePrefix = "arXiv",
    primaryClass = "hep-ex",
    doi = "10.1103/PhysRevD.110.012009",
    journal = "Phys. Rev. D",
    volume = "110",
    number = "1",
    pages = "012009",
    year = "2024"
}

@article{SNDLHC:2022ihg,
    author = "Acampora, G. and others",
    collaboration = "SND@LHC",
    title = "{SND@LHC: the scattering and neutrino detector at the LHC}",
    eprint = "2210.02784",
    archivePrefix = "arXiv",
    primaryClass = "hep-ex",
    doi = "10.1088/1748-0221/19/05/P05067",
    journal = "JINST",
    volume = "19",
    number = "05",
    pages = "P05067",
    year = "2024"
}

@article{FASER:2022hcn,
    author = "Abreu, Henso and others",
    collaboration = "FASER",
    title = "{The FASER detector}",
    eprint = "2207.11427",
    archivePrefix = "arXiv",
    primaryClass = "physics.ins-det",
    reportNumber = "CERN-FASER-2022-001",
    doi = "10.1088/1748-0221/19/05/P05066",
    journal = "JINST",
    volume = "19",
    number = "05",
    pages = "P05066",
    year = "2024"
}

@article{SHiP:2020sos,
    author = "Ahdida, C. and others",
    collaboration = "SHiP",
    title = "{SND@LHC}",
    eprint = "2002.08722",
    archivePrefix = "arXiv",
    primaryClass = "physics.ins-det",
    reportNumber = "CERN-LHCC-2020-002, LHCC-I-035",
    month = "2",
    year = "2020"
}

@inproceedings{Roesler:2000he,
    author = "Roesler, Stefan and Engel, Ralph and Ranft, Johannes",
    title = "{The Monte Carlo event generator DPMJET-III}",
    booktitle = "{International Conference on Advanced Monte Carlo for Radiation Physics, Particle Transport Simulation and Applications (MC 2000)}",
    eprint = "hep-ph/0012252",
    archivePrefix = "arXiv",
    reportNumber = "SLAC-PUB-8740",
    doi = "10.1007/978-3-642-18211-2_166",
    pages = "1033--1038",
    month = "12",
    year = "2000"
}

@article{Ballarini:2024isa,
    author = "Ballarini, Francesca and others",
    title = "{The FLUKA code: Overview and new developments}",
    doi = "10.1051/epjn/2024015",
    journal = "EPJ Nuclear Sci. Technol.",
    volume = "10",
    pages = "16",
    year = "2024"
}

@article{DsTau:2025,
    author = "Aoki, S. and others",
    collaboration = "DsTau/NA65",
    title = "{Study of proton--nucleus interactions in the DsTau/NA65 experiment at the CERN-SPS}",
    eprint = "2411.05452",
    archivePrefix = "arXiv",
    primaryClass = "hep-ex",
    doi = "10.1140/epjc/s10052-025-13928-1",
    journal = "Eur. Phys. J. C",
    volume = "85",
    number = "3",
    pages = "230",
    year = "2025"
}

@article{Ariga:2025gtj,
    author = {Ariga, Akitaka and Barwick, Steven and Boyd, Jamie and Fieg, Max and Kling, Felix and M{\"a}kel{\"a}, Toni and Vendeuvre, Camille and Weyer, Benjamin},
    title = "{Detecting LHC neutrinos at surface level}",
    eprint = "2501.06142",
    archivePrefix = "arXiv",
    primaryClass = "hep-ex",
    doi = "10.1007/JHEP07(2025)270",
    journal = "JHEP",
    volume = "07",
    pages = "270",
    year = "2025"
}

@article{IceCube:2021rpz,
    author = "Aartsen, M. G. and others",
    collaboration = "IceCube",
    title = "{Detection of a particle shower at the Glashow resonance with IceCube}",
    eprint = "2110.15051",
    archivePrefix = "arXiv",
    primaryClass = "hep-ex",
    doi = "10.1038/s41586-021-03256-1",
    journal = "Nature",
    volume = "591",
    number = "7849",
    pages = "220--224",
    year = "2021",
    note = "[Erratum: Nature 592, E11 (2021)]"
}

@article{Blednykh:2026gii,
    author = "Blednykh, Alexei",
    title = "{Collective effects: Challenges and solutions for the EIC project}",
    doi = "10.18429/JACoW-NAPAC2025-THYD01",
    journal = "JACoW",
    volume = "NAPAC2025",
    pages = "THYD01",
    year = "2026"
}

@misc{GarciaSoto:2025shift,
      title = "{A SHIFT of Perspective: Observing Neutrinos at CMS and ATLAS}",
      author = "Garcia-Soto, Alfonso and Niedziela, Jeremi",
      year = "2025",
      eprint = "2510.11816",
      archivePrefix = "arXiv",
      primaryClass = "hep-ph"
}

@article{Deshpande:2002em,
    author = "Deshpande, A. L.",
    editor = "Bass, S. D. and De Roeck, A. and Deshpande, A.",
    title = "{The EIC project: Physics prospects and present status}",
    doi = "10.1016/S0920-5632(01)01977-6",
    journal = "Nucl. Phys. B Proc. Suppl.",
    volume = "105",
    pages = "178--184",
    year = "2002"
}

@article{FASER:2025sge,
    author = "Mammen Abraham, Roshan and others",
    collaboration = "FASER",
    title = "{Latest neutrino results from the FASER experiment and their implications for forward hadron production}",
    eprint = "2507.23552",
    archivePrefix = "arXiv",
    primaryClass = "hep-ex",
    reportNumber = "CERN-FASER-CONF-2025-004",
    doi = "10.22323/1.501.0349",
    journal = "PoS",
    volume = "ICRC2025",
    pages = "349",
    year = "2025"
}

@article{KM3NeT:2025npi,
    author = "Aiello, S. and others",
    collaboration = "KM3NeT",
    title = "{Observation of an ultra-high-energy cosmic neutrino with KM3NeT}",
    doi = "10.1038/s41586-024-08543-1",
    journal = "Nature",
    volume = "638",
    number = "8050",
    pages = "376--382",
    year = "2025",
    note = "[Erratum: Nature 640, E3 (2025)]"
}

@article{Andreopoulos:2009rq,
    author = "Andreopoulos, C. and others",
    title = "{The GENIE Neutrino Monte Carlo Generator}",
    eprint = "0905.2517",
    archivePrefix = "arXiv",
    primaryClass = "hep-ph",
    reportNumber = "FERMILAB-PUB-09-418-CD",
    doi = "10.1016/j.nima.2009.12.009",
    journal = "Nucl. Instrum. Meth. A",
    volume = "614",
    pages = "87--104",
    year = "2010"
}

@article{Andreopoulos:2015wxa,
    author = "Andreopoulos, Costas and Barry, Christopher and Dytman, Steve and Gallagher, Hugh and Golan, Tomasz and Hatcher, Robert and Perdue, Gabriel and Yarba, Julia",
    title = "{The GENIE Neutrino Monte Carlo Generator: Physics and User Manual}",
    eprint = "1510.05494",
    archivePrefix = "arXiv",
    primaryClass = "hep-ph",
    reportNumber = "FERMILAB-FN-1004-CD",
    month = "10",
    year = "2015"
}

@article{AbdulKhalek:2021gbh,
    author = "Abdul Khalek, Rabah and others",
    title = "{Science Requirements and Detector Concepts for the Electron-Ion Collider: {EIC} Yellow Report}",
    eprint = "2103.05419",
    archivePrefix = "arXiv",
    primaryClass = "physics.ins-det",
    doi = "10.1016/j.nuclphysa.2022.122447",
    journal = "Nucl. Phys. A",
    volume = "1026",
    pages = "122447",
    year = "2022"
}

@article{Kling:2021gos,
    author = "Kling, Felix and Nevay, Laurence J.",
    title = "{Forward neutrino fluxes at the LHC}",
    eprint = "2105.08270",
    archivePrefix = "arXiv",
    primaryClass = "hep-ph",
    doi = "10.1103/PhysRevD.104.113008",
    journal = "Phys. Rev. D",
    volume = "104",
    number = "11",
    pages = "113008",
    year = "2021"
}

@article{SNDLHC:2023pun,
    author = "Albanese, R. and others",
    collaboration = "SND@LHC",
    title = "{Observation of Collider Muon Neutrinos with the SND@LHC Experiment}",
    eprint = "2305.09383",
    archivePrefix = "arXiv",
    primaryClass = "hep-ex",
    reportNumber = "CERN-EP-2023-092",
    doi = "10.1103/PhysRevLett.131.031802",
    journal = "Phys. Rev. Lett.",
    volume = "131",
    number = "3",
    pages = "031802",
    year = "2023"
}

@article{Piparo:2023yam,
    author = "Piparo, G. and others",
    title = "{Measurement of the forward {\ensuremath{\eta}} meson production rate in p-p collisions at $ \sqrt{\textrm{s}} $ = 13 TeV with the LHCf-Arm2 detector}",
    eprint = "2305.06633",
    archivePrefix = "arXiv",
    primaryClass = "hep-ex",
    reportNumber = "CERN-EP-2023-076",
    doi = "10.1007/JHEP10(2023)169",
    journal = "JHEP",
    volume = "10",
    pages = "169",
    year = "2023"
}

@article{LHCf:2020hjf,
    author = "Adriani, O. and others",
    collaboration = "LHCf",
    title = "{Measurement of energy flow, cross section and average inelasticity of forward neutrons produced in $ \sqrt{s} $ = 13 TeV proton-proton collisions with the LHCf Arm2 detector}",
    eprint = "2003.02192",
    archivePrefix = "arXiv",
    primaryClass = "hep-ex",
    reportNumber = "CERN-EP-2020-029",
    doi = "10.1007/JHEP07(2020)016",
    journal = "JHEP",
    volume = "07",
    pages = "016",
    year = "2020"
}

@article{FASER:2024hoe,
    collaboration = "FASER",
    title = "{First Measurement of \ensuremath{\nu_e} and \ensuremath{\nu_\mu} Interaction Cross Sections at the LHC with FASER\textquoteright{}s Emulsion Detector}",
    eprint = "2403.12520",
    archivePrefix = "arXiv",
    primaryClass = "hep-ex",
    reportNumber = "CERN-EP-2024-079",
    doi = "10.1103/PhysRevLett.133.021802",
    journal = "Phys. Rev. Lett.",
    volume = "133",
    number = "2",
    pages = "021802",
    year = "2024"
}

@article{FASER:2018eoc,
    collaboration = "FASER",
    title = "{FASER\textquoteright{}s physics reach for long-lived particles}",
    eprint = "1811.12522",
    archivePrefix = "arXiv",
    primaryClass = "hep-ph",
    reportNumber = "UCI-TR-2018-19, KYUSHU-RCAPP-2018-06",
    doi = "10.1103/PhysRevD.99.095011",
    journal = "Phys. Rev. D",
    volume = "99",
    number = "9",
    pages = "095011",
    year = "2019"
}

@article{FASER:2023tle,
    collaboration = "FASER",
    title = "{Search for dark photons with the FASER detector at the LHC}",
    eprint = "2308.05587",
    archivePrefix = "arXiv",
    primaryClass = "hep-ex",
    reportNumber = "CERN-EP-2023-161",
    doi = "10.1016/j.physletb.2023.138378",
    journal = "Phys. Lett. B",
    volume = "848",
    pages = "138378",
    year = "2024"
}

@misc{SNDMoriond26,
    author = "S. Ilieva",
    collaboration = "SND@LHC",
    howpublished = {\textit{Recent results from the SND@LHC experiment [Conference presentation]}, Rencontres de Moriond QCD and high-energy interactions, \url{https://moriond.in2p3.fr/2026/QCD/}},
    year = "2026"
}

@article{FASER:2023zcr,
    author = "Abreu, Henso and others",
    collaboration = "FASER",
    title = "{First Direct Observation of Collider Neutrinos with FASER at the LHC}",
    eprint = "2303.14185",
    archivePrefix = "arXiv",
    primaryClass = "hep-ex",
    reportNumber = "CERN-EP-2023-056",
    doi = "10.1103/PhysRevLett.131.031801",
    journal = "Phys. Rev. Lett.",
    volume = "131",
    number = "3",
    pages = "031801",
    year = "2023"
}

@article{FASER:2018bac,
    author = "Ariga, Akitaka and others",
    collaboration = "FASER",
    title = "{Technical Proposal for FASER: ForwArd Search ExpeRiment at the LHC}",
    eprint = "1812.09139",
    archivePrefix = "arXiv",
    primaryClass = "physics.ins-det",
    reportNumber = "CERN-LHCC-2018-036, LHCC-P-013, UCI-TR-2018-22, KYUSHU-RCAPP-2018-07",
    month = "12",
    year = "2018"
}

@article{Feng:2017uoz,
    author = "Feng, Jonathan L. and Galon, Iftah and Kling, Felix and Trojanowski, Sebastian",
    title = "{ForwArd Search ExpeRiment at the LHC}",
    eprint = "1708.09389",
    archivePrefix = "arXiv",
    primaryClass = "hep-ph",
    reportNumber = "UCI-TR-2017-08",
    doi = "10.1103/PhysRevD.97.035001",
    journal = "Phys. Rev. D",
    volume = "97",
    number = "3",
    pages = "035001",
    year = "2018"
}

@article{FASER:2021mtu,
    author = "Abreu, Henso and others",
    collaboration = "FASER",
    title = "{First neutrino interaction candidates at the LHC}",
    eprint = "2105.06197",
    archivePrefix = "arXiv",
    primaryClass = "hep-ex",
    doi = "10.1103/PhysRevD.104.L091101",
    journal = "Phys. Rev. D",
    volume = "104",
    number = "9",
    pages = "L091101",
    year = "2021"
}

@article{FASER:2019dxq,
    author = "Abreu, Henso and others",
    collaboration = "FASER",
    title = "{Detecting and Studying High-Energy Collider Neutrinos with FASER at the LHC}",
    eprint = "1908.02310",
    archivePrefix = "arXiv",
    primaryClass = "hep-ex",
    reportNumber = "CERN-EP-2019-160, KYUSHU-RCAPP-2019-003, SLAC-PUB-17460, UCI-TR-2019-19, SLAC-PUB-17460,
  UCI-TR-2019-19",
    doi = "10.1140/epjc/s10052-020-7631-5",
    journal = "Eur. Phys. J. C",
    volume = "80",
    number = "1",
    pages = "61",
    year = "2020"
}

@article{Foldenauer:2021gkm,
    author = "Foldenauer, Patrick and Kling, Felix and Reimitz, Peter",
    title = "{Potential of CMS as a high-energy neutrino scattering experiment}",
    eprint = "2108.05370",
    archivePrefix = "arXiv",
    primaryClass = "hep-ph",
    reportNumber = "IPPP/21/10, DESY-21-120",
    doi = "10.1103/PhysRevD.104.113005",
    journal = "Phys. Rev. D",
    volume = "104",
    number = "11",
    pages = "113005",
    year = "2021"
}

@article{Beni:2019gxv,
    author = "Beni, N. and others",
    title = "{Physics Potential of an Experiment using LHC Neutrinos}",
    eprint = "1903.06564",
    archivePrefix = "arXiv",
    primaryClass = "hep-ex",
    reportNumber = "CMS-NOTE-2019-001",
    doi = "10.1088/1361-6471/ab3f7c",
    journal = "J. Phys. G",
    volume = "46",
    number = "11",
    pages = "115008",
    year = "2019"
}

@article{XSEN:2019bel,
    author = "Beni, N. and others",
    collaboration = "XSEN",
    title = "{XSEN: a $\nu$N Cross Section Measurement using High Energy Neutrinos from pp collisions at the LHC}",
    eprint = "1910.11340",
    archivePrefix = "arXiv",
    primaryClass = "physics.ins-det",
    reportNumber = "CERN-LHCC-2019-014 / LHCC-I-033",
    month = "10",
    year = "2019"
}

@article{DeRujula:1992sn,
      author         = "De Rujula, A. and Fernandez, E. and Gomez-Cadenas, J. J.",
      title          = "{Neutrino fluxes at future hadron colliders}",
      journal        = "Nucl. Phys. B",
      volume         = "405",
      pages          = "80--108",
      year           = "1993",
      doi            = "10.1016/0550-3213(93)90427-Q",
      reportNumber   = "CERN-TH-6452-92, IFAE-92-001",
      SLACcitation   = "%%CITATION = NUPHA,B405,80;%%"
}

@article{Park:2011gh,
      author         = "Park, HyangKyu",
      title          = "{The estimation of neutrino fluxes produced by proton-proton collisions at $\sqrt{s}=14$ TeV of the LHC}",
      journal        = "JHEP",
      volume         = "10",
      year           = "2011",
      pages          = "092",
      doi            = "10.1007/JHEP10(2011)092",
      eprint         = "1110.1971",
      archivePrefix  = "arXiv",
      primaryClass   = "hep-ex",
      SLACcitation   = "%%CITATION = ARXIV:1110.1971;%%"
}

@article{Anchordoqui:2021ghd,
    author = "Anchordoqui, Luis A. and others",
    title = "{The Forward Physics Facility: Sites, experiments, and physics potential}",
    eprint = "2109.10905",
    archivePrefix = "arXiv",
    primaryClass = "hep-ph",
    reportNumber = "BNL-222142-2021-FORE, CERN-PBC-Notes-2021-025, DESY-21-142, DESY-21-142,
  FERMILAB-CONF-21-452-AE-E-ND-PPD-T, KYUSHU-RCAPP-2021-01, LU TP 21-36,
  PITT-PACC-2118, SMU-HEP-21-10, UCI-TR-2021-22, FERMILAB-CONF-21-452-AE-E-ND-PPD-T",
    doi = "10.1016/j.physrep.2022.04.004",
    journal = "Phys. Rept.",
    volume = "968",
    pages = "1--50",
    year = "2022"
}

@article{Feng:2022inv,
    author = "Feng, Jonathan L. and others",
    title = "{The Forward Physics Facility at the High-Luminosity LHC}",
    eprint = "2203.05090",
    archivePrefix = "arXiv",
    primaryClass = "hep-ex",
    reportNumber = "UCI-TR-2022-01, CERN-PBC-Notes-2022-001, INT-PUB-22-006, BONN-TH-2022-04, FERMILAB-PUB-22-094-ND-SCD-T",
    doi = "10.1088/1361-6471/ac865e",
    journal = "J. Phys. G",
    volume = "50",
    number = "3",
    pages = "030501",
    year = "2023"
}

@article{Worcester:2023njy,
    author = "Worcester, Elizabeth",
    title = "{The Dawn of Collider Neutrino Physics}",
    doi = "10.1103/Physics.16.113",
    journal = "APS Physics",
    volume = "16",
    pages = "113",
    month = "7",
    year = "2023"
}

@article{Abraham:2024fcc,
    author = "Abraham, Roshan Mammen and Adhikary, Jyotismita and Feng, Jonathan L. and Fieg, Max and Kling, Felix and Li, Jinmian and Pei, Junle and Rabemananjara, Tanjona R. and Rojo, Juan and Trojanowski, Sebastian",
    title = "{FPF@FCC: Neutrino, QCD, and BSM Physics Opportunities with Far-Forward Experiments at a 100 TeV Proton Collider}",
    eprint = "2409.02163",
    archivePrefix = "arXiv",
    primaryClass = "hep-ph",
    doi = "10.1007/JHEP01(2025)094",
    journal = "JHEP",
    volume = "01",
    pages = "094",
    year = "2025"
}

@article{Kling:2026NSI,
    author = "Kling, Felix and Ma, Yang and M{\k e}ka{\l}a, Krzysztof and Reuter, J{\"u}rgen and Tabrizi, Zahra",
    title = "{Non-standard neutrino interactions at a muon collider neutrino detector}",
    eprint = "2508.00761",
    archivePrefix = "arXiv",
    primaryClass = "hep-ph",
    doi = "10.1007/JHEP05(2026)050",
    journal = "JHEP",
    volume = "05",
    pages = "050",
    year = "2026"
}

@article{Marzocca:2025qmk,
    author = "Marzocca, David and Montagno, Francesco and Morales-Alvarado, Manuel and Wulzer, Andrea",
    title = "{Quark mixing from muon collider neutrinos}",
    eprint = "2511.23288",
    archivePrefix = "arXiv",
    primaryClass = "hep-ph",
    doi = "10.1007/JHEP05(2026)069",
    journal = "JHEP",
    volume = "05",
    pages = "069",
    year = "2026"
}

@article{Adhikary:2025npd,
    author = "Adhikary, Jyotismita and Kelly, Kevin J. and Kling, Felix and Trojanowski, Sebastian",
    title = "{Neutrino-portal dark matter detection prospects at a future muon collider}",
    eprint = "2412.10315",
    archivePrefix = "arXiv",
    primaryClass = "hep-ph",
    doi = "10.1103/PhysRevD.111.075019",
    journal = "Phys. Rev. D",
    volume = "111",
    number = "7",
    pages = "075019",
    year = "2025"
}

@article{CruzMartinez:2023sdv,
    author = {Cruz-Martinez, Juan M. and Fieg, Max and Giani, Tommaso and Krack, Peter and M\"akel\"a, Toni and Rabemananjara, Tanjona R. and Rojo, Juan},
    title = "{The LHC as a Neutrino-Ion Collider}",
    eprint = "2309.09581",
    archivePrefix = "arXiv",
    primaryClass = "hep-ph",
    reportNumber = "Nikhef-2023-009, CERN-TH-2023-165",
    doi = "10.1140/epjc/s10052-024-12665-1",
    journal = "Eur. Phys. J. C",
    volume = "84",
    number = "4",
    pages = "369",
    year = "2024"
}

@article{Boyarsky:2021moj,
    author = "Boyarsky, Alexey and Mikulenko, Oleksii and Ovchynnikov, Maksym and Shchutska, Lesya",
    title = "{Searches for new physics at SND@LHC}",
    eprint = "2104.09688",
    archivePrefix = "arXiv",
    primaryClass = "hep-ph",
    doi = "10.1007/JHEP03(2022)006",
    journal = "JHEP",
    volume = "03",
    pages = "006",
    year = "2022"
}

@article{FASER:2024bbl,
    author = "Mammen Abraham, Roshan and others",
    collaboration = "FASER",
    title = "{Shining Light on the Dark Sector: Search for Axion-like Particles and Other New Physics in Photonic Final States with FASER}",
    eprint = "2410.10363",
    archivePrefix = "arXiv",
    primaryClass = "hep-ex",
    reportNumber = "CERN-EP-2024-262",
    doi = "10.1007/JHEP01(2025)199",
    journal = "JHEP",
    volume = "01",
    pages = "199",
    year = "2025"
}

@article{Pierog:2013ria,
    author = "Pierog, T. and Karpenko, Iu. and Katzy, J. M. and Yatsenko, E. and Werner, K.",
    title = "{EPOS LHC: Test of collective hadronization with data measured at the CERN Large Hadron Collider}",
    eprint = "1306.0121",
    archivePrefix = "arXiv",
    primaryClass = "hep-ph",
    reportNumber = "DESY-13-125",
    doi = "10.1103/PhysRevC.92.034906",
    journal = "Phys. Rev. C",
    volume = "92",
    number = "3",
    pages = "034906",
    year = "2015"
}

@article{Ostapchenko:2010vb,
    author = "Ostapchenko, Sergey",
    title = "{Monte Carlo treatment of hadronic interactions in enhanced Pomeron scheme: I. QGSJET-II model}",
    eprint = "1010.1869",
    archivePrefix = "arXiv",
    primaryClass = "hep-ph",
    doi = "10.1103/PhysRevD.83.014018",
    journal = "Phys. Rev. D",
    volume = "83",
    pages = "014018",
    year = "2011"
}

@article{Riehn:2019jet,
    author = "Riehn, Felix and Engel, Ralph and Fedynitch, Anatoli and Gaisser, Thomas K. and Stanev, Todor",
    title = "{Hadronic interaction model Sibyll 2.3d and extensive air showers}",
    eprint = "1912.03300",
    archivePrefix = "arXiv",
    primaryClass = "hep-ph",
    doi = "10.1103/PhysRevD.102.063002",
    journal = "Phys. Rev. D",
    volume = "102",
    number = "6",
    pages = "063002",
    year = "2020"
}

@article{Fieg:2023kld,
    author = {Fieg, Max and Kling, Felix and Schulz, Holger and Sj\"ostrand, Torbj\"orn},
    title = "{Tuning pythia for forward physics experiments}",
    eprint = "2309.08604",
    archivePrefix = "arXiv",
    primaryClass = "hep-ph",
    reportNumber = "DESY-23-133",
    doi = "10.1103/PhysRevD.109.016010",
    journal = "Phys. Rev. D",
    volume = "109",
    number = "1",
    pages = "016010",
    year = "2024"
}

@article{Alioli:2010xd,
    author = "Alioli, Simone and Nason, Paolo and Oleari, Carlo and Re, Emanuele",
    title = "{A general framework for implementing NLO calculations in shower Monte Carlo programs: the POWHEG BOX}",
    eprint = "1002.2581",
    archivePrefix = "arXiv",
    primaryClass = "hep-ph",
    reportNumber = "DESY-10-018, SFB-CPP-10-22, IPPP-10-11, DCPT-10-22",
    doi = "10.1007/JHEP06(2010)043",
    journal = "JHEP",
    volume = "06",
    pages = "043",
    year = "2010"
}

@article{Bierlich:2022pfr,
    author = "Bierlich, Christian and others",
    title = "{A comprehensive guide to the physics and usage of PYTHIA 8.3}",
    eprint = "2203.11601",
    archivePrefix = "arXiv",
    primaryClass = "hep-ph",
    reportNumber = "LU-TP 22-16, MCNET-22-04, FERMILAB-PUB-22-227-SCD",
    doi = "10.21468/SciPostPhysCodeb.8",
    journal = "SciPost Phys. Codeb.",
    volume = "2022",
    pages = "8",
    year = "2022"
}

@article{Bodek:2002vp,
    author = "Bodek, A. and Yang, U. K.",
    editor = "Morfin, J. G. and Sakuda, M. and Suzuki, Y.",
    title = "{Modeling deep inelastic cross-sections in the few GeV region}",
    eprint = "hep-ex/0203009",
    archivePrefix = "arXiv",
    doi = "10.1016/S0920-5632(02)01755-3",
    journal = "Nucl. Phys. B Proc. Suppl.",
    volume = "112",
    pages = "70--76",
    year = "2002"
}

@article{Bodek:2004pc,
    author = "Bodek, Arie and Park, Inkyu and Yang, Un-ki",
    editor = "Cavanna, F. and Keppel, C. and Lipari, P. and Sakuda, M.",
    title = "{Improved low Q2 model for neutrino and electron nucleon cross sections in few GeV region}",
    eprint = "hep-ph/0411202",
    archivePrefix = "arXiv",
    doi = "10.1016/j.nuclphysbps.2004.11.208",
    journal = "Nucl. Phys. B Proc. Suppl.",
    volume = "139",
    pages = "113--118",
    year = "2005"
}

@article{Bodek:2010km,
    author = "Bodek, Arie and Yang, Un-ki",
    title = "{Axial and Vector Structure Functions for Electron- and Neutrino- Nucleon Scattering Cross Sections at all $Q^2$ using Effective Leading order Parton Distribution Functions}",
    eprint = "1011.6592",
    archivePrefix = "arXiv",
    primaryClass = "hep-ph",
    month = "11",
    year = "2010"
}

@article{FASER:2024ref,
    author = "Mammen Abraham, Roshan and others",
    collaboration = "FASER",
    title = "{First Measurement of the Muon Neutrino Interaction Cross Section and Flux as a Function of Energy at the LHC with FASER}",
    eprint = "2412.03186",
    archivePrefix = "arXiv",
    primaryClass = "hep-ex",
    reportNumber = "CERN-EP-2024-309",
    doi = "10.1103/PhysRevLett.134.211801",
    journal = "Phys. Rev. Lett.",
    volume = "134",
    number = "21",
    pages = "211801",
    year = "2025"
}

@article{SNDLHC:2024qqb,
    author = "Abbaneo, D. and others",
    collaboration = "SND@LHC",
    title = "{Observation of Collider Neutrinos without Final State Muons with the SND@LHC Experiment}",
    eprint = "2411.18787",
    archivePrefix = "arXiv",
    primaryClass = "hep-ex",
    reportNumber = "CERN-EP-2024-316",
    doi = "10.1103/r2qy-9hft",
    journal = "Phys. Rev. Lett.",
    volume = "134",
    number = "23",
    pages = "231802",
    year = "2025"
}

@article{Brock:1993sz,
  author = {Brock, R. and others},
  title = {Handbook of perturbative {QCD}},
  journal = {Rev. Mod. Phys.},
  volume = {67},
  year = {1995},
  pages = {157},
  doi = {10.1103/RevModPhys.67.157}
}

@article{Conrad:1997ne,
  author = {Conrad, J. and Shaevitz, M. and Bolton, T.},
  title = {Precision measurements with high-energy neutrino beams},
  journal = {Rev. Mod. Phys.},
  volume = {70},
  year = {1998},
  pages = {1341},
  doi = {10.1103/RevModPhys.70.1341},
  eprint = "hep-ex/9707015",
  archivePrefix = "arXiv",
  primaryClass = "hep-ex",
}

@article{Formaggio:2013kya,
  author = {Formaggio, J. A. and Zeller, G. P.},
  title = {From {eV} to {EeV}: Neutrino cross sections across energy scales},
  journal = {Rev. Mod. Phys.},
  volume = {84},
  year = {2012},
  pages = {1307},
  doi = {10.1103/RevModPhys.84.1307},
  eprint = "1305.7513",
  archivePrefix = "arXiv",
  primaryClass = "hep-ex",
}

@article{DeLellis:2004sg,
  author = {De Lellis, G. and Migliozzi, P. and Santorelli, P.},
  title = {Charm physics with neutrinos},
  journal = {Phys. Rept.},
  volume = {399},
  year = {2004},
  pages = {227},
  doi = {10.1016/j.physrep.2004.07.005}
}

@inproceedings{DeRujula:1984pg,
    author = "De Rujula, A. and Ruckl, R.",
    title = "{Neutrino and muon physics in the collider mode of future accelerators}",
    booktitle = "{SSC Workshop: Superconducting Super Collider Fixed Target Physics}",
    reportNumber = "CERN-TH-3892/84",
    doi = "10.5170/CERN-1984-010-V-2.571",
    pages = "571--596",
    month = "5",
    year = "1984"
}

@article{Marfatia:2015hja,
  author = {Marfatia, D. and McKay, D. and Weiler, T.},
  title = {New physics with ultra-high-energy neutrinos},
  journal = {Phys. Lett. B},
  volume = {748},
  year = {2015},
  pages = {113},
  doi = {10.1016/j.physletb.2015.07.002},
  eprint = "1502.06337",
  archivePrefix = "arXiv",
  primaryClass = "hep-ph",
}

@article{Arguelles:2019ziu,
  author = {Arg{\"u}elles, C. A. and others},
  title = {New opportunities at the next-generation neutrino experiments {I}: {BSM} neutrino physics and dark matter},
  journal = {Rept. Prog. Phys.},
  volume = {83},
  year = {2020},
  pages = {124201},
  doi = {10.1088/1361-6633/ab9d12},
  eprint = "1907.08311",
  archivePrefix = "arXiv",
  primaryClass = "hep-ph",
}

@article{Aad:2008zzm,
    author = "Aad, G. and others",
    collaboration = "ATLAS",
    title = "{The ATLAS Experiment at the CERN Large Hadron Collider}",
    doi = "10.1088/1748-0221/3/08/S08003",
    journal = "JINST",
    volume = "3",
    pages = "S08003",
    year = "2008"
}

@article{Chatrchyan:2008zzk,
    author = "Chatrchyan, S. and others",
    collaboration = "CMS",
    title = "{The CMS Experiment at the CERN LHC}",
    doi = "10.1088/1748-0221/3/08/S08004",
    journal = "JINST",
    volume = "3",
    pages = "S08004",
    year = "2008"
}

@article{Aamodt:2008zz,
    author = "Aamodt, K. and others",
    collaboration = "ALICE",
    title = "{The ALICE experiment at the CERN LHC}",
    doi = "10.1088/1748-0221/3/08/S08002",
    journal = "JINST",
    volume = "3",
    pages = "S08002",
    year = "2008"
}

@article{Alves:2008zz,
    author = "Alves, Jr., A. Augusto and others",
    collaboration = "LHCb",
    title = "{The LHCb Detector at the LHC}",
    reportNumber = "LHCb-DP-2008-001",
    doi = "10.1088/1748-0221/3/08/S08005",
    journal = "JINST",
    volume = "3",
    pages = "S08005",
    year = "2008"
}

@article{CooperSarkar:2011pa,
  author = {Cooper-Sarkar, A. and Mertsch, P. and Sarkar, S.},
  title = {The high energy neutrino cross-section in the {Standard Model} and its uncertainty},
  journal = {JHEP},
  volume = {08},
  year = {2011},
  pages = {042},
  doi = {10.1007/JHEP08(2011)042},
  eprint = "1106.3723",
  archivePrefix = "arXiv",
  primaryClass = "hep-ph",
}

@article{Graverini:2024ynx,
  author = {Graverini, E.},
  title = {A roadmap for neutrino detection at {LHC}, {HL-LHC} and {SPS}},
  journal = {Nucl. Instrum. Meth. A},
  volume = {1068},
  year = {2024},
  pages = {169804},
  doi = {10.1016/j.nima.2024.169804},
  note = {on behalf of the SND@LHC and SHiP Collaborations},
  eprint = "2408.15851",
  archivePrefix = "arXiv",
  primaryClass = "hep-ex",
}

@article{Adhikary:2024nlv,
  author = {Adhikary, J. and others},
  collaboration = {FPF Working Groups},
  title = {Scientific program for the {Forward Physics Facility}},
  journal = {Eur. Phys. J. C},
  volume = {85},
  year = {2025},
  pages = {430},
  doi = {10.1140/epjc/s10052-025-14048-6},
  eprint = {2411.04175},
  archivePrefix = {arXiv},
  primaryClass = {hep-ex}
}

@article{Kopp:2007ne,
    author = "Kopp, Joachim and Lindner, Manfred",
    title = "{Detecting atmospheric neutrino oscillations in the ATLAS detector at CERN}",
    eprint = "0705.2595",
    archivePrefix = "arXiv",
    primaryClass = "hep-ph",
    doi = "10.1103/PhysRevD.76.093003",
    journal = "Phys. Rev. D",
    volume = "76",
    pages = "093003",
    year = "2007"
}

@article{Bustamante:2017xuy,
    author = "Bustamante, Mauricio and Connolly, Amy",
    title = "{Extracting the Energy-Dependent Neutrino-Nucleon Cross Section above 10 TeV Using IceCube Showers}",
    eprint = "1711.11043",
    archivePrefix = "arXiv",
    primaryClass = "astro-ph.HE",
    doi = "10.1103/PhysRevLett.122.041101",
    journal = "Phys. Rev. Lett.",
    volume = "122",
    number = "4",
    pages = "041101",
    year = "2019"
}

@article{Kamp:2025phs,
    author = {Kamp, Nicholas W. and Arg{\"u}elles, Carlos A. and Karle, Albrecht and Thomas, Jennifer and Yuan, Tianlu},
    title = "{Lake- and surface-based detectors for forward neutrino physics}",
    eprint = "2501.08278",
    archivePrefix = "arXiv",
    primaryClass = "hep-ex",
    doi = "10.1103/z4f4-wdc3",
    journal = "Phys. Rev. D",
    volume = "113",
    number = "5",
    pages = "052002",
    year = "2026"
}

@article{Ariga:2025qup,
    author = "Ariga, Akitaka and Boyd, Jamie and Kling, Felix and De Roeck, Albert",
    title = "{Neutrino Experiments at the Large Hadron Collider}",
    eprint = "2501.10078",
    archivePrefix = "arXiv",
    primaryClass = "hep-ex",
    doi = "10.1146/annurev-nucl-121423-101000",
    journal = "Ann. Rev. Nucl. Part. Sci.",
    volume = "75",
    number = "1",
    pages = "57--81",
    year = "2025"
}

@inproceedings{Zhang:2026cpk,
    author = "Zhang, Shunliang and Hu, Zhen",
    title = "{Latest Results from the FASER Experiment}",
    booktitle = "{60th Rencontres de Moriond on QCD and High Energy Interactions}: {Moriond QCD 2026}",
    eprint = "2604.16244",
    archivePrefix = "arXiv",
    primaryClass = "hep-ex",
    month = "4",
    year = "2026"
}

@article{FASER:2025myb,
    author = "Mammen Abraham, Roshan and others",
    collaboration = "FASER",
    title = "{Prospects and Opportunities with an upgraded FASER Neutrino Detector during the HL-LHC era: Input to the EPPSU}",
    eprint = "2503.19775",
    archivePrefix = "arXiv",
    primaryClass = "hep-ex",
    reportNumber = "CERN-FASER-2025-001",
    month = "3",
    year = "2025"
}

@article{SNDLHC:2026why,
    author = "Abbaneo, D. and others",
    collaboration = "SND@LHC",
    title = "{SND@LHC Upgrade for the High-Luminosity LHC: Physics Reach and Installation Scenarios}",
    eprint = "2602.21881",
    archivePrefix = "arXiv",
    primaryClass = "hep-ex",
    month = "2",
    year = "2026"
}

@article{Wen:2023ijf,
    author = {Wen, Alex Y. and Arg{\"u}elles, Carlos A. and Kheirandish, Ali and Murase, Kohta},
    title = "{Detecting High-Energy Neutrinos from Galactic Supernovae with ATLAS}",
    eprint = "2309.09771",
    archivePrefix = "arXiv",
    primaryClass = "hep-ph",
    doi = "10.1103/PhysRevLett.132.061001",
    journal = "Phys. Rev. Lett.",
    volume = "132",
    number = "6",
    pages = "061001",
    year = "2024"
}

@article{Ghosh:2024ryg,
    author = "Ghosh, Deep and Mukhopadhyay, Satyanarayan and Mukhopadhyaya, Biswarup",
    title = "{Prospects of measuring the atmospheric muon neutrino and antineutrino flux ratio with the ATLAS detector}",
    eprint = "2409.20231",
    archivePrefix = "arXiv",
    primaryClass = "hep-ph",
    doi = "10.1103/pj4c-cn41",
    journal = "Phys. Rev. D",
    volume = "113",
    pages = "112009",
    year = "2026"
}

@techreport{ESPP:2020,
    author = "{European Strategy Group}",
    title = "{2020 Update of the European Strategy for Particle Physics (Brochure)}",
    institution = "CERN",
    address = "Geneva, Switzerland",
    number = "CERN-ESU-015",
    year = "2020",
    url = "https://cds.cern.ch/record/2721370"
}

@techreport{ESPP:2026Recommendations,
    author = "{The European Strategy Group (ESG)}",
    title = "{The European Strategy for Particle Physics: 2026 Update - Recommendations by the European Strategy Group}",
    institution = "CERN",
    address = "Geneva, Switzerland",
    number = "CERN-ESU-2025-002",
    year = "2025",
    url = "https://cds.cern.ch/record/2950671"
}

@techreport{Abbaneo:SNDHLHC:LHCC-P-026,
    author = "Abbaneo, D. and others",
    title = "{SND@HL-LHC, Scattering and Neutrino Detector in Run 4 of the LHC}",
    institution = "CERN",
    address = "Geneva, Switzerland",
    number = "CERN-LHCC-2025-004; LHCC-P-026",
    year = "2025",
    url = "https://cds.cern.ch/record/2926288"
}

@techreport{Salin:FASER2:PBCNOTE2025-005,
    author = "Salin, Olivier and others",
    title = "{FASER2: Detector Design and Performance}",
    institution = "CERN",
    address = "Geneva, Switzerland",
    number = "CERN-PBC-Notes-2025-005",
    year = "2025",
    url = "https://cds.cern.ch/record/2927003"
}

@techreport{Conf-FASER-CONF-2025-001,
    collaboration = "FASER",
    title = "{Measurement of the Muon Neutrino Flux as a Function of Rapidity with FASER}",
    institution = "CERN",
    address = "Geneva, Switzerland",
    number = "CERN-FASER-CONF-2025-001",
    year = "2025",
    url = "https://cds.cern.ch/record/2927629"
}

@techreport{Conf-FASER-CONF-2026-001,
    collaboration = "FASER",
    title = "{New Results from a Search for Dark Photons with the FASER Detector at the LHC}",
    institution = "CERN",
    address = "Geneva, Switzerland",
    number = "CERN-FASER-CONF-2026-001",
    year = "2026",
    url = "https://cds.cern.ch/record/2955719?ln=en"
}

@techreport{Conf-FASER-CONF-2026-002,
    collaboration = "FASER",
    title = "{Cross section measurements of high-energy electron and muon neutrino interactions with FASER's emulsion detector at the LHC}",
    institution = "CERN",
    address = "Geneva, Switzerland",
    number = "CERN-FASER-CONF-2026-002",
    year = "2026",
    url = "https://cds.cern.ch/record/2956166?ln=en"
}

@techreport{Conf-FASER-CONF-2026-003,
    collaboration = "FASER",
    title = "{First search for neutrino-induced charm hadrons with FASER's emulsion detector at the LHC}",
    institution = "CERN",
    address = "Geneva, Switzerland",
    number = "CERN-FASER-CONF-2026-003",
    year = "2026",
    url = "https://cds.cern.ch/record/2956186?ln=en"
}

@techreport{Conf-FASER-CONF-2026-004,
    collaboration = "FASER",
    title = "{Measurement of High-Energy Electron Neutrino Interactions with the FASER Calorimeter at the LHC}",
    institution = "CERN",
    address = "Geneva, Switzerland",
    number = "CERN-FASER-CONF-2026-004",
    year = "2026",
    url = "https://cds.cern.ch/record/2956447?ln=en"
}

@techreport{Conf-FASER-CONF-2026-005,
    collaboration = "FASER",
    title = "{Measurement of Muon Neutrinos as a Function of Energy and Rapidity with FASER}",
    institution = "CERN",
    address = "Geneva, Switzerland",
    number = "CERN-FASER-CONF-2026-005",
    year = "2026",
    url = "https://cds.cern.ch/record/2956617?ln=en"
}

@article{Anchordoqui:2025fpf,
    author = "Anchordoqui, Luis A. and others",
    collaboration = "FPF",
    title = "{The Forward Physics Facility: Physics Opportunities and Conceptual Design}",
    eprint = "2510.26260",
    archivePrefix = "arXiv",
    primaryClass = "hep-ex",
    reportNumber = "CERN-PBC-Notes-2025-010",
    doi = "10.1016/j.nuclphysb.2026.117398",
    journal = "Nucl. Phys. B",
    volume = "1026",
    pages = "117398",
    year = "2026"
}

@article{milliQan:2021lne,
    author = "Ball, A. and others",
    collaboration = "milliQan",
    title = "{Sensitivity to millicharged particles in future proton-proton collisions at the LHC with the milliQan detector}",
    eprint = "2104.07151",
    archivePrefix = "arXiv",
    primaryClass = "hep-ex",
    doi = "10.1103/PhysRevD.104.032002",
    journal = "Phys. Rev. D",
    volume = "104",
    number = "3",
    pages = "032002",
    year = "2021"
}

@article{SHiP:2021nfo,
    author = "Ahdida, C. and others",
    collaboration = "SHiP",
    title = "{The SHiP experiment at the proposed CERN SPS Beam Dump Facility}",
    eprint = "2112.01487",
    archivePrefix = "arXiv",
    primaryClass = "physics.ins-det",
    doi = "10.1140/epjc/s10052-022-10346-5",
    journal = "Eur. Phys. J. C",
    volume = "82",
    number = "5",
    pages = "486",
    year = "2022"
}

@article{Boyd:2026xuf,
    author = "Boyd, Jamie",
    title = "{The FASER experiment at the Large Hadron Collider}",
    eprint = "2604.19199",
    archivePrefix = "arXiv",
    primaryClass = "hep-ex",
    month = "4",
    year = "2026"
}

@article{SNDLHC:2026muf,
    author = "Abbaneo, D. and others",
    collaboration = "SND@LHC",
    title = "{Measurement of the Muon Flux at SND@LHC: Results from the 2023--2025 Proton and Heavy-Ion Periods}",
    eprint = "2602.23412",
    archivePrefix = "arXiv",
    primaryClass = "hep-ex",
    reportNumber = "CERN-EP-2026-039",
    month = "2",
    year = "2026"
}

@article{SNDLHC:2026yqa,
    author = "Abbaneo, D. and others",
    collaboration = "SND@LHC",
    title = "{Study of the Run-3 muon flux at the SND@LHC experiment}",
    eprint = "2603.21878",
    archivePrefix = "arXiv",
    primaryClass = "hep-ex",
    reportNumber = "CERN-EP-2026-086",
    doi = "10.1140/epjc/s10052-026-15875-x",
    journal = "Eur. Phys. J. C",
    volume = "86",
    pages = "732",
    year = "2026"
}

@article{FASER:2025reco,
    author = "Mammen Abraham, Roshan and others",
    collaboration = "FASER",
    title = "{Reconstruction and Performance Evaluation of FASER's Emulsion Detector at the LHC}",
    eprint = "2504.13008",
    archivePrefix = "arXiv",
    primaryClass = "physics.ins-det",
    doi = "10.1088/1748-0221/20/12/P12018",
    journal = "JINST",
    volume = "20",
    pages = "P12018",
    year = "2025"
}

@article{FASER:2026mcs,
    collaboration = "FASER",
    title = "{Momentum Measurement of Charged Particles in FASER's Emulsion Detector at the LHC}",
    eprint = "2602.17575",
    archivePrefix = "arXiv",
    primaryClass = "physics.ins-det",
    year = "2026"
}

@article{SNDLHC:2026oxu,
    author = "Abbaneo, D. and others",
    collaboration = "SND@LHC",
    title = "{Measurement of the muon neutrino charged-current cross section with SND@LHC}",
    eprint = "2606.14669",
    archivePrefix = "arXiv",
    primaryClass = "hep-ex",
    month = "6",
    year = "2026"
}

@misc{SNDICHEP26,
    author = "C. Asawatangtrakuldee",
    collaboration = "SND@LHC",
    howpublished = {\textit{Recent results from the Scattering and Neutrino Detector at the LHC (SND@LHC) [Conference presentation]}, 43rd International Conference on High Energy Physics (ICHEP 2026), \url{https://indico.cern.ch/event/1522800/contributions/6985922/}},
    year = "2026"
}

@book{BejarAlonso:2020hup,
    editor = "Béjar Alonso, I. and Brüning, O. and Fessia, P. and Lamont, M. and Rossi, L. and Tavian, L. and Zerlauth, M.",
    title = "{High-Luminosity Large Hadron Collider (HL-LHC): Technical Design Report}",
    series = "CERN Yellow Reports: Monographs",
    volume = "10",
    reportNumber = "CERN-2020-010",
    publisher = "CERN",
    address = "Geneva",
    doi = "10.23731/CYRM-2020-0010",
    year = "2020"
}

\end{document}